\documentclass[12pt]{article}

\usepackage{graphicx}
\usepackage{epsf}

\usepackage{amsmath,amssymb,float} 

\usepackage{pstricks}
\usepackage{color}
\usepackage{multirow}
\usepackage{cite}

\def\beq{\begin{equation}}
\def\eq{\end{equation}}
\def\eeq{\end{equation}}

\def\centeron#1#2{{\setbox0=\hbox{#1}\setbox1=\hbox{#2}\ifdim
\wd1>\wd0\kern.5\wd1\kern-.5\wd0\fi
\copy0\kern-.5\wd0\kern-.5\wd1\copy1\ifdim\wd0>\wd1
\kern.5\wd0\kern-.5\wd1\fi}}
\def\ltap{\;\centeron{\raise.35ex\hbox{$<$}}{\lower.65ex\hbox{$\sim$}}\;}
\def\gtap{\;\centeron{\raise.35ex\hbox{$>$}}{\lower.65ex\hbox{$\sim$}}\;}
\def\gsim{\mathrel{\gtap}}

\def\chii0{\chi_i^0}
\def\chij0{\chi_j^0}

\def\foursqr#1#2{{\vcenter{\vbox{
 \hrule height.#2pt
 \hbox{\vrule width.#2pt height#1pt \kern#1pt
 \vrule width.#2pt}
 \hrule height.#2pt
 \hrule height.#2pt
 \hbox{\vrule width.#2pt height#1pt \kern#1pt
 \vrule width.#2pt}
 \hrule height.#2pt
     \hrule height.#2pt
 \hbox{\vrule width.#2pt height#1pt \kern#1pt
 \vrule width.#2pt}
 \hrule height.#2pt
     \hrule height.#2pt
 \hbox{\vrule width.#2pt height#1pt \kern#1pt
 \vrule width.#2pt}
 \hrule height.#2pt}}}}
\def\psqr#1#2{{\vcenter{\vbox{\hrule height.#2pt
 \hbox{\vrule width.#2pt height#1pt \kern#1pt
 \vrule width.#2pt}
 \hrule height.#2pt \hrule height.#2pt
 \hbox{\vrule width.#2pt height#1pt \kern#1pt
 \vrule width.#2pt}
 \hrule height.#2pt}}}}
\def\sqr#1#2{{\vcenter{\vbox{\hrule height.#2pt
 \hbox{\vrule width.#2pt height#1pt \kern#1pt
 \vrule width.#2pt}
 \hrule height.#2pt}}}}

\def\figin{\epsfcheck\figin}\def\figins{\epsfcheck\figins}
\def\epsfcheck{\ifx\epsfbox\UnDeFiNeD
\message{(NO epsf.tex, FIGURES WILL BE IGNORED)}
\gdef\figin##1{\vskip2in}\gdef\figins##1{\hskip.5in}
\else\message{(FIGURES WILL BE INCLUDED)}%
\gdef\figin##1{##1}\gdef\figins##1{##1}\fi}
\def\DefWarn#1{}
\def\figinsert{\goodbreak\midinsert}
\def\ifig#1#2#3{\DefWarn#1\xdef#1{fig.~\the\figno}
\writedef{#1\leftbracket fig.\noexpand~\the\figno}%
\figinsert\figin{\centerline{#3}}\medskip\centerline{\vbox{\baselineskip12pt
\advance\hsize by -1truein\noindent\footnotefont{\bf
Fig.~\the\figno:\ } \it#2}}
\bigskip\endinsert\global\advance\figno by1}

\def\fig#1#2#3#4{\vskip 0.5cm \begingroup \midinsert \centerline{
\psfig{file=#1,width=#2}} \vskip 0.4cm
\global\advance\figno by 1
\centerline{\vbox{\baselineskip=12pt \noindent Figure \the\figno: #3}}
\endinsert \endgroup {\xdef#4{\the\figno}} }

\def\figcrop#1#2#3#4#5#6#7#8{\vskip 0.5cm \begingroup \midinsert \centerline{
\psfig{file=#1,width=#2,bbllx=#3,bblly=#4,bburx=#5,bbury=#6}} \vskip 0.4cm
\global\advance\figno by 1
\centerline{\vbox{\baselineskip=12pt \noindent Figure \the\figno: #7}}
\endinsert \endgroup {\xdef#8{\the\figno}} \vskip .5cm}

\def\figlabel#1{\xdef#1{\the\figno}}
\def\encadremath#1{\vbox{\hrule\hbox{\vrule\kern8pt\vbox{\kern8pt
\hbox{$\displaystyle #1$}\kern8pt}
\kern8pt\vrule}\hrule}}
\def\underarrow#1{\vbox{\ialign{##\crcr$\hfil\displaystyle
 {#1}\hfil$\crcr\noalign{\kern1pt\nointerlineskip}$\longrightarrow$\crcr}}}

\begin{document}

\begin{titlepage}

\begin{center}
\vspace*{-1cm}

\vskip 0.7in
{\LARGE \bf Signals of New Resonances from Di-Lepton  } \\
\vspace{.12in}
{\LARGE \bf Non-Universality in the Bottomonium  } \\
\vspace{.12in}
{\LARGE\bf Mass Region at the Large Hadron Collider } \\
\vspace{.12in}

\vskip 0.55in
{\large Connor Houghton},~
{\large Amit Lath},~ 
{\large Joseph Reichert},~
{\large Scott Thomas}

\vskip 0.2in
{\em 
Department of Physics, 
Rutgers University, 
Piscataway, NJ 08854}

\vskip 0.7in

\end{center}

\baselineskip=16pt


\noindent
Universality classes of new physics 
models featuring narrow boson resonances produced through strongly interacting initial 
 states in proton-proton collisions and with enhanced decays to di-tau final states are classified.  
Spin-zero scalar or pseudoscalar bosons with chirality-violating fermion couplings
 can induce potentially significant 
 di-lepton \protect \linebreak
  non-universality in the bottomonium mass region, with up to many hundreds of nanobarns 
 of direct production total cross section in proton-proton collisions at the Large Hadron Collider. 
Conversely, mass suppressed chirality-violating 
 couplings of spin-zero bosons to di-electrons maintain di-lepton universality to a high degree in the same mass 
region at electron-positron colliders. 
Simultaneous measurement and comparison of integrated resonant prompt di-electron, di-muon, and di-tau 
mass spectra at the Large Hadron Collider in the bottomonium mass region could be sensitive to 
the existence of new boson resonances 
with enhanced di-tau decays, or likewise bottomonium states with non-Standard Model 
enhanced di-tau decay modes.


\end{titlepage}

\baselineskip=17pt

\newpage


\section{Introduction}

The search for 
new physics at the 
Large Hadron Collider (LHC) encompasses an extensive range of signals over a vast landscape of experimental signatures. 
A prime focus of these efforts is the direct production in proton-proton collisions of new resonant states that 
  decay into detectable final states. 
 Among these, di-lepton channels 
   can provide relatively clean probes for new 
   di-electron \cite{CMS:2019lwf, CMS:2024ulc, CMS:2021ctt, ATLAS:2019erb}, 
 di-muon 
 \cite{CMS:2019lwf, CMS:2024ulc,CMS:2012fgd,CMS:2019buh,CMS:2021sch,CMS:2023hwl,LHCb:2017trq,LHCb:2019vmc,LHCb:2020ysn}, 
 and di-tau 
 \cite{CMS:2024ulc, CMS:2014ccx, CMS:2015qnd, CMS:2018rmh, CMS:2022goy, ATLAS:2014vhc, ATLAS:2016ivh,
   ATLAS:2017eiz, ATLAS:2020zms, ATLAS:2024rzd} 
 resonances.   
 Compared with searches for resonant hadronic final states, which can be obscured by significant quantum chromodynamic (QCD) backgrounds, 
 resonant di-lepton signatures are characterized by relatively manageable continuum 
 backgrounds. 

Resonant di-lepton searches in mass regions near or overlapping known Standard Model (SM)
 resonances require meticulous background characterization. 
For the heavy resonances of the Higgs and $Z$ bosons,  first-principles SM perturbative Monte Carlo simulations 
provide a reliable 
framework for modeling hard-scattering production cross sections and decay kinematics. 
However, for low mass narrow di-lepton resonances, including 
$\Upsilon, J/ \psi, \phi, \omega$ and $\eta$, no such simulations are yet available that can fully and accurately predict
from first principles the SM hard-scattering non-perturbative production of these composite bound states. 
To circumvent the complexities and uncertainties in 
modeling the production of these resonant composite states, as well as others that can contribute continuum di-lepton 
backgrounds through decays, 
standard search strategies typically 
employ parametric fits to the di-lepton continuum along with blinded mass windows centered on 
the narrow composite di-lepton resonances. 
This has become standard practice in low mass di-muon resonance searches \cite{LHCb:2020ysn}. 
While this approach avoids the requirement of rigorous theoretical calculations of SM backgrounds, 
it introduces a significant limitation: 
 sensitivity is drastically reduced or entirely eliminated for any new di-lepton resonance of mass near or within the blinded regions. 

In this paper we propose to circumvent the inherent limitations of searches for new di-lepton resonances near known 
 composite resonances, specifically in the bottomonium mass region. 
This is the only region with known composite resonances that  decay directly to all three 
generations of di-electron, di-muon, and di-tau final states with significant branching ratio.   
Direct decays to prompt di-leptons from the myriad of bottomonium states that can be produced at the LHC
are dominated by the $\Upsilon(nS)$ resonances with $n=1,2,3$. 
These decay through off-shell photons and are di-lepton universal in the SM at the sub-percent level. 
Other bottomonium states have negligible direct decays to prompt di-lepton final states. 
The Drell-Yan continuum background is also di-lepton universal at a similar level.  
So SM production of prompt di-leptons at a hadron collider is expected to yield di-lepton universality 
 to a high degree throughout the entire bottomonium mass region. 

The di-muon mass spectrum in the bottomonium mass region, with prominent $\Upsilon(nS)$ resolved resonances, 
 is well measured in proton-proton collisions at the LHC 
 \cite{ATLAS:2012lmu,ATLAS:2017prf,LHCb:2012aa,LHCb:2018yzj,CMS:2013qur,CMS:2015xqv,CMS:2017dju,CMS:2026ccg,Zhao:2025sna}. 
Observation and measurement of di-electron \cite{CMS:2024zhe} and di-tau \cite{CMS:2026mwx}    
 mass spectra 
 in the same conditions 
 has recently 
 been made possible through the use of low-threshold high band-width triggers 
  with compressed event-level information.  
The $\Upsilon(nS)$ resonances are distinguishable in the di-electron mass spectrum \cite{CMS:2024zhe} albeit with
 somewhat degraded mass resolution compared with di-muon. 
 The resolution in the di-tau visible 
 mass spectrum reconstructed from coincident muonic and hadronic decays of di-tau 
  \cite{CMS:2026mwx}  
  is not sufficient to resolve individual $\Upsilon(nS)$ resonances, 
 but essentially integrates over all contributions within the bottomonium mass region.  
In all three cases the resonant contributions to the individual di-lepton mass spectra above the continuum 
 background in the bottomonium mass region can now be observed and measured. 
 
The proposal is to make a simultaneous measurement of 
the resonant contributions to the prompt  
di-electron, di-muon, and di-tau mass spectra integrated over the bottomonium mass region. 
 SM bottomonium decays are expected to give essentially universal contributions
to each category.  
A measurement such as this could be sensitive to 
new boson resonances with non-universal di-lepton decays
that lie within the bottomonium mass region. 
It could also be sensitive to non-SM decays of bottomonium states to non-universal di-lepton decay modes.  
Since di-muon has the best resolution and reconstruction efficiency, as a practical matter this would amount to comparing 
the integrated di-electron and/or di-tau resonant spectra with the resonant di-muon spectrum. 
And since di-tau has the least well measured mass resolution, it could be sensitive to either a new boson resonance
with mass anywhere within the bottomonium mass range 
with enhanced di-tau decays, 
 or likewise bottomonium states with enhanced di-tau decay modes. 

Ideally, measurement and comparison of resonant prompt di-lepton mass spectra 
in the bottomonium mass region 
  would be carried out within a single analysis using a common trigger. 
A single muon trigger would be optimal in this regard.  
It would capture di-tau through muonic decays of one tau-lepton 
 and allow reconstruction of hadronic decays of the second.  
 It would also capture di-muon, where the kinematics accepted for the second muon could be adjusted 
  to on average closely match that of the hadronically decaying tau-lepton, thereby 
   yielding similar kinematic acceptance across di-tau and di-muon channels. 
A single electron trigger could likewise be employed for di-tau and di-electron channels, 
although with some reduced energy resolution, 
 and larger fake-rate backgrounds.

Universality classes of new physics models that feature fairly narrow boson resonances 
produced through strongly interacting initial states and with decays that can directly contribute to 
 di-lepton non-universality in the bottomonium mass region
   are presented 
    in the sections below. 
 Enhanced decay branching ratios to di-tau final states over di-muon and di-electron  
 arise naturally for spin-zero bosons 
 with chirality-violating couplings to di-leptons. 
  %
Total cross sections times branching ratio to di-tau final states can be up to many hundreds of 
 nanobarns, and after acceptances in appropriate kinematic fiducial regions many 
  nanobarns.  
 Sub-classes of ultraviolet completions for the origin of interactions of spin-zero scalar and pseudoscalar 
  bosons with 
  SM di-fermions, including from mixing with (extended) Higgs sector states and heavy vector-like fermions 
 are also illustrated.  
    These universality classes may be used to parameterize and interpret 
    results of 
  measurements 
 of di-lepton (non)-universality in the bottomonium mass region at the LHC.  
 Similar models of spin-zero pseudoscalar bosons with mixing through an extended Higgs sector 
 have been considered recently \cite{Buckley:2026xcv}.

Tests for di-lepton non-universality and  enhanced di-tau final states in $\Upsilon(nS)$  production 
 and decays have been carried out at electron-positron colliders \cite{CLEO:2006uhx, BaBar:2010esv, BaBar:2020nlq} 
  with all results consistent with di-lepton universality.  
 The measurements proposed here to test for di-lepton non-universality in the bottomonium mass region 
  in proton-proton collisions at the LHC are complementary. 
 The  spin-zero boson resonances of the new physics universality classes considered below have highly suppressed 
  di-electron couplings 
with small direct production cross sections at electron-positron colliders.
 In addition, a myriad of bottomonium states, beyond 
  the $\Upsilon(nS)$ accessible through direct production at an electron-positron,
 are directly produced in proton-proton collisions, and any of these with enhanced di-tau decay modes
 could contribute to di-lepton non-universality in the measurements proposed here at the LHC. 
 Some of these additional states, as well as the new physics resonances considered here, 
may only accessible at a electron-positron collider through 
  cascade decays of $\Upsilon(nS)$.



\vspace{0.1in}

\section {Di-Lepton Non-Universality from New Bosons \\ in the Bottomonium Mass Region } 

General universality classes of new physics models that can induce di-lepton 
 non-universality in the bottomonium mass region in proton-proton collisions 
 are outlined in the 
 sections below. 
Attention is focused on classes of models with relatively narrow boson resonances 
 that can be produced in the $s$-channel through strongly interacting initial states and with 
  enhanced decay branching ratios to di-tau final states. 
  Only direct production contributions of a new physics boson to di-lepton non-universality 
 are considered here.

In order to organize the classification and presentation, a number of simplifying restrictions are made.  
Only the leading most relevant interactions of the bosons with SM fermions and gauge bosons are presented 
 in the vacuum with 
 massive electroweak gauge bosons. 
Interactions of the bosons in these classes with the Higgs boson are also interesting 
and relevant to other signals not considered here.
The flavor structure of 
interactions of the bosons with leptons and quarks are restricted to commute with, or equivalently 
 align with, the flavor 
 structure of the SM lepton and quark Yukawa couplings respectively. 
 In boson interactions this enforces that first, individual lepton flavor is conserved, 
 and second that quark sector flavor-changing neutral current interactions are absent at tree-level. 
 Finally, all interactions of the bosons are taken to respect time-reversal symmetry. 
 This is particularly relevant for spin-zero bosons 
that may be classified in this case as scalar or pseudoscalar, with even and odd transformation properties
 under time-reversal respectively. 
 Of course, any or all of these simplifying restrictions may or may not be relaxed in matching to particular 
 ultraviolet completions, but they do provide well defined universality classes of models. 
 
 
 For the Monte Carlo (MC) simulations presented below,  gluon fusion production of spin-zero scalar and pseudoscalar boson
  events are generated at next to leading order (NLO) in QCD interactions with the gg\textsc{\_H\_2HDM} module 
  implemented in \textsc{Powheg Box}
   v2~\cite{Frixione:2007vw,Alioli:2010xd,Bagnaschi:2011tu}
 using the \textsc{NNPDF31}\_nlo\_as\_0118 parton distribution functions~\cite{NNPDF:2017mvq}.
 Depending on the scenario simulated,  
top and bottom quark loop contributions, or a heavy quark loop limit, are included in the gluon fusion production processes. 
 For scalar and pseudoscalar boson production in association with a bottom and anti-bottom quark, events are generated 
  in four-flavor scheme to leading order (LO) in QCD interactions utilizing the SM plugin of 
 \textsc{MadGraph5\_aMC@NLO} v3.7.0~\cite{Alwall:2014hca} 
  including up to one additional final state parton, with the MLM jet matching scheme~\cite{Alwall:2007fs} 
  using the four-flavor \textsc{NNPDF30}\_lo\_as\_0118\_nf\_4 parton distribution functions~\cite{NNPDF:2014otw}.
 In the case of scalar boson production
  in association with a bottom and anti-bottom quark, K-factors are obtained for these LO simulations by comparing with 
 events generated at NLO in QCD interactions 
 using the {loop\_sm} plugin with up to one additional final state parton, 
  applying the FxFx jet matching scheme~\cite{Frederix:2012ps} using the \textsc{NNPDF31}\_nlo\_nf\_4 parton distribution  functions~\cite{NNPDF:2017mvq}.
  These K-factors are applied to the LO pseudoscalar boson production in association with a bottom and anti-bottom quark. 
All scalar and pseudoscalar spin-zero boson production events are showered utilizing Pythia 8.313 \cite{Bierlich:2022pfr}. 
 Proton-proton initial states are used in all cases with 13.6 TeV center of mass energy. 
The produced boson masses in all cases are taken to be 10 GeV, 
roughly in the center of the relevant bottomonium mass region. 

Kinematic acceptances are calculated directly for the bosons in the MC simulations. 
No final state detector or reconstruction efficiencies are applied.  
Benchmark values of kinematic requirements for the bosons used to calculate the 
fiducial kinematic acceptances reported below are taken to be transverse momentum 
$p_T > 20$ GeV and rapidity $|y| < 1.2$. 
These 
roughly approximate the effective 
fiducial kinematic 
acceptance of visible di-tau decay products in 
recent observation and measurement of resonant contributions to
 di-tau final states 
 in the bottomonium mass region \cite{CMS:2026mwx}. 

Branching ratios for spin-zero 
bosons are calculated using the 
partial decay widths given in Appendix A. 
The di-gluon partial widths include top, bottom, and charm quark loops, 
and the di-photon partial width includes these as well the tau-lepton loop. 
For two-body fermion partial widths only the charm quark and tau-lepton final states are included. 
For the benchmark boson mass of 10 GeV considered here, there are no two-body final 
 states that include a bottom quark in one state and  anti-bottom quark in the other. 
 It is worth noting that for a similar reason, the perturbative 
one-loop bottom quark contribution to MC simulation of gluon fusion production, 
as well as calculation of the di-gluon partial width, receive non-trivial non-perturbative corrections 
 for boson masses in the bottomonium mass range.  
In particular, using the bottom quark pole mass for the loop kinematics with the benchmark boson mass here, 
the perturbative one-loop bottom quark loops have an absorptive component for on-shell intermediate states, 
whereas in reality this is absent for physical confined bound states.  
In the MC simulations and branching ratio calculations, 
 the strong and electromagnetic fine structure constants, as well 
  as the $\overline{\rm MS}$ quark masses used are for a renormalization scale equal to the boson mass of 10 GeV, 
 with the exception of the top quark for which a renormalization scale equal to the top quark mass is used.



\section{Spin-Zero Scalar or Pseudoscalar Bosons}

The leading interactions in a momentum expansion of spin-zero bosons to di-fermions may be written, 
 up to operator relations, as Yukawa-type couplings. 
In this representation the interactions are fermion chirality-violating, 
 coupling left-handed to right-handed chiral components of the fermions. 
 
In the SM the only renormalizable 
 marginal chirality-violating fermion interactions are Yukawa couplings with the Higgs field. 
And since these Yukawa interactions are responsible for fermion masses in the vacuum with 
massive electroweak gauge bosons, 
 all chirality-violating fermion interactions in the SM are proportional to fermion masses.   
If individual (discrete) chiral symmetry 
is restored as each lepton SM Yukawa coupling is taken to zero in a full theory that includes additional
 spin-zero bosons, then the leading couplings of the new bosons to di-fermions
 also vanish in these limits, and are proportional to fermion masses. 
 Such interactions may be classified as mass-proportional chirality-violating fermion non-universality,  
  possibly with flavor dependent proportionality.

The chirality-violating nature of di-fermion couplings to spin-zero bosons furnishes a natural 
 source of di-lepton non-universality in the decay of these  bosons. 
 This is realized for exact mass-proportional couplings, or with approximate possibly flavor dependent 
  proportionality. 
   The general universality classes of spin-zero new physics models considered below are restricted to ones with 
  such exact or approximate mass-proportional couplings to di-fermions.  
    The specific interactions and resulting  
    proton-proton collision direct production cross sections and decay branching
  ratios 
in these classes that can contribute 
 to lepton non-universality in the bottomonium mass region are detailed below. 
 
  Scalar and pseudoscalar spin-zero bosons have different allowed couplings
 under the restriction of time-reversal symmetry. 
 However, many of the general features 
  and magnitude of potential contributions to di-lepton non-universality in the bottomonium 
   mass region are similar for both cases. 
  The quantitative direct production contributions from either a scalar or pseudoscalar spin-zero boson 
   are presented as
  separate independent scenarios.

Sub-classes of some ultraviolet completions that can give rise to the spin-zero boson
 interactions in these general universality classes are  presented in the sub-section below,     
including quark and lepton couplings inherited through mixings from (extended) 
 Higgs sector Yukawa couplings, or from couplings to heavy vector-like 
  quarks and leptons that mix with SM quark and leptons, 
   as well as gluon couplings coming from interactions with heavy vector-like quarks.



The allowed 
relevant and marginal interactions, up to operator relations, that are 
linear in a single real spin-zero scalar boson $\phi$ and couple to SM Dirac fermions $\psi$ and SM 
spin-one gauge bosons
in the vacuum with massive electroweak bosons, are 
\beq 
 - \, g_{\phi \psi \psi} \, {m_\psi \over v} \, \, \phi \, \, \bar{\psi}   \,  \psi
  + 2 \, g_{\phi ZZ} \, \phi \, \, {m_Z^2 \over v} \, Z_\mu Z^\mu 
  + 2 \, g_{\phi WW} \,  \phi \, {m_W^2  \over v} \, W^+_\mu W^{-\mu}
  \label{scalarcouplings} 
\eq
where by definition $\phi=0$ in the vacuum. 
The mass-proportional nature of couplings to di-fermions within the universality classes considered here, 
 is illustrated  with the explicit factor of fermion mass $m_\psi$. 
The dimensionless coefficients $g_{\phi \psi \psi}$ parameterize the possibly flavor dependent 
 constants of proportionality. 
The normalizing mass-scale factor in both the di-fermion and gauge field couplings is taken to be the SM Higgs 
doublet field 
 expectation value $v = \sqrt{2} \, \langle H^0 \rangle \simeq 246$ GeV. 
 For reference, in the case where the scalar boson is the Higgs boson, $\phi = h$, 
all the dimensionless coupling coefficients are unity 
$g_{h \psi \psi} = g_{h ZZ} = g_{h WW}= 1$. 
This normalization is particularly convenient for making contact with the 
sub-classes of ultraviolet completions presented below in which the couplings of the scalar  boson 
 are inherited from mixing with (extended) Higgs sector states. 
 Only the scalar couplings to di-fermions in (\ref{scalarcouplings}) are of direct relevance to 
 di-lepton non-universality 
  in the bottomonium mass region. 
   The magnitude of the allowed couplings to massive spin-one SM gauge bosons in (\ref{scalarcouplings}) 
   are considered in 
  discussion of the sub-classes of ultraviolet completions below.

The only allowed marginal interactions, up to operator relations, that are 
linear in a single real spin-zero pseudoscalar boson $\varpi$, often referred to  
 as an axion-like particle,  in the vacuum with massive electroweak bosons,
  are couplings to SM Dirac fermions $\psi$
\beq 
 - \, g_{\varpi \psi \psi} \, {m_\psi \over v} \, \, \varpi \, \, \bar{\psi} \, i \gamma_5  \,  \psi
 \label{pseudoscalarcouplings} 
\eq
where by definition $\varpi = 0$ in the vacuum.
Time-reversal symmetry forbids any marginal interactions with SM gauge fields.   
The mass-proportional nature of couplings to di-fermions within the universality classes considered here, 
 is illustrated  with the explicit factor of fermion mass $m_\psi$, with the same normalizing mass-scale factor  
 of the SM Higgs field expectation value $v$ used in the scalar interactions (\ref{scalarcouplings}). 
 The dimensionless coefficients $g_{\varpi \psi \psi}$ parameterize the possibly flavor dependent 
 constants of proportionality.  
This normalization is also convenient for making contact with the 
sub-classes of ultraviolet completions presented below in which the couplings of the pseudoscalar  boson 
 are inherited from mixing with extended Higgs sector states.

The 
 direct production cross section 
in proton-proton collisions
 for a spin-zero scalar or pseudoscalar boson coming from the marginal di-fermion interactions 
(\ref{scalarcouplings}) 
 or (\ref{pseudoscalarcouplings}) 
 depends on the dimensionless coupling coefficients $g_{\phi \psi \psi}$ or $g_{\varpi \psi \psi}$.
For definiteness, and to simplify the discussion, 
 the dimensionless coupling coefficients for all up-type quarks are taken to be 
 limited  from above by roughly unity.  
 Likewise, the coupling coefficients for all down-type quarks are taken to be limited 
  from above by roughly $m_t / m_b$. 
  For the third generation top and bottom quarks this ensures that Yukawa couplings to the 
  spin-zero boson do not parametrically exceed their infrared quasi-fixed point values. 
 With these restrictions the dominant production mode of a spin-zero scalar or pseudoscalar 
  boson from strongly interacting initial states 
  in proton-proton collisions is gluon fusion through top and bottom quark loops. 
  Production of a spin-zero boson in association with a bottom quark and anti-bottom quark 
   is smaller by roughly an order of magnitude than the bottom quark loop contribution 
    to gluon fusion, and is also included below.  
  
The decay branching ratios of a spin-zero scalar or pseudoscalar boson with 
marginal di-fermion interactions 
(\ref{scalarcouplings}) 
 or (\ref{pseudoscalarcouplings}) also 
 depend on the dimensionless coupling coefficients $g_{\phi \psi \psi}$ or $g_{\varpi \psi \psi}$.
As long as the lepton chirality-violating di-fermion couplings are approximately 
mass-proportional, with dimensionless coupling coefficients not differing significantly in magnitude  
 among the three lepton flavors,  the di-tau final state dominates the di-lepton decay modes. 
 As commented above, for the benchmark boson mass of 10 GeV, 
 there are no two or more body final states with a bottom quark in one state and an anti-bottom 
 quark in another.  
 So with the parametric restrictions on dimensional coupling coefficients specified above, 
  decays to strongly interacting final states  are then dominated at the parton level by di-charm 
  or di-gluon coming predominantly from top, bottom, and charm quark loops.

In order to illustrate the quantitative contributions of a spin-zero scalar or pseudoscalar boson to 
di-lepton non-universality in the bottomonium mass region, 
 it is instructive to consider some discrete choices for the 
 dimensionless coupling coefficients $g_{\phi \psi \psi}$ or $g_{\varpi \psi \psi}$.  
 The results for the di-tau branching ratio 
 ${\rm Br}( \phi, \varpi \to \tau \tau)$,  
 the total inclusive cross section 
 $\sigma(pp \to \phi, \varpi) $,
  and cross section times branching ratio times acceptance in the benchmark fiducial kinematic region 
 $\sigma(pp \to \phi , \varpi) \cdot  {\rm Br}( \phi, \varpi \to \tau \tau) \, \cdot 
    {\rm Acc}(p_{T}^{\phi, \varpi } \! > \! 20 \, {\rm GeV} \, , \, | y_{\phi, \varpi} | < 1.2) $, 
  are presented for 
three limiting patterns of non-vanishing coupling coefficients 
in Table 1  for scalar or pseudoscalar bosons of mass 10 GeV.


\begin{table}[p]  
\begin{center}  \vspace*{0.2in} 
\begin{tabular}{cccc} 
  Non-Zero  & ${\rm Br}( \phi , \varpi \to \tau \tau)$  & $\sigma(pp \to \phi , \varpi ) $ & 
  $\sigma(pp \to \phi , \varpi ) \cdot  {\rm Br}( \phi , \varpi  \to \tau \tau) \, \cdot $   \\
Couplings  & & [nb]&    ${\rm Acc}(p_{T}^{\phi , \varpi } \! > \! 20 \, {\rm GeV} \, , \, | y_{\phi  , \varpi} | < 1.2) $    \\
  & & & [nb] \\
  & \\
 $g_{\phi \tau \tau } = g_{\phi tt}$  &  0.99  &  1.06 $g_{\phi tt}^2  $  & 0.095 $g_{\phi tt}^2  $   \\
& \\ 
 $g_{\phi \tau \tau } =  g_{\phi bb}$  & 0.99  & 1.72 $g_{\phi bb}^2 $  & 0.017 $g_{\phi bb}^2 $      \\
  & \\
 $ g_{\phi \psi \psi} \, \, \forall \, \, \psi $ &  0.56   & 4.57 $g_{\phi \psi \psi}^2 $    &  0.088 $g_{\phi \psi \psi}^2 $    \\
 & \\
 $ g_{\phi \tau \tau} \, , \, M_{\phi gg} $  &   
   $ 0.99$ 
    &  1.03 $(v / M_{\phi gg})^2 $  
       &   0.092   $(v / M_{\phi gg})^2 $    
       \\
 & \\
  & \\
 $g_{\varpi \tau \tau } = g_{\varpi tt}$  &  0.98  &  1.62 $g_{\varpi tt}^2  $  & 0.150 $g_{\varpi tt}^2  $   \\
& \\ 
 $g_{\varpi \tau \tau } =  g_{\varpi bb}$  & 0.95  & 1.35 $g_{\varpi bb}^2 $  & 0.018 $g_{\varpi bb}^2 $      \\
  & \\
 $ g_{\varpi \psi \psi} \, \, \forall \, \, \psi $ &  0.54   & 4.41 $g_{\varpi \psi \psi}^2 $    &  0.113 $g_{\varpi \psi \psi}^2 $    \\
 & \\
 $ g_{\varpi \tau \tau} \, , \, M_{\varpi gg} $  &   
   $ 0.98$ 
   &  1.55 $(v / M_{\varpi gg})^2 $  
       &   0.143   $(v / M_{\varpi gg})^2 $    
        \\
 & \\
  \end{tabular}
\caption{Di-tau branching ratio ${\rm Br}( \phi , \varpi \to \tau \tau)$, 
 total inclusive cross section $\sigma(pp \to \phi , \varpi) $ in nanobarns 
    from gluon fusion and where appropriate production in association 
     with a bottom and anti-bottom quark, and total inclusive production cross section 
        times di-tau branching ratio times acceptance  
   $\sigma(pp \to \phi , \varpi  ) \cdot {\rm Br}( \phi , \varpi \to \tau \tau) 
    \cdot {\rm Acc}(p_{T}^{\phi , \varpi } \! > \! 20 \, {\rm GeV} \, , \, | y_{\phi , \varpi } | < 1.2) 
  $
       of a spin-zero scalar $\phi$ or pseudoscalar $\varpi$ of mass $m_{\phi , \varpi } = 10$ GeV 
     in 13.6 TeV proton-proton collisions for various 
     combinations of chirality-violating di-tau and di-quark dimensionless coupling coefficients
     $g_{\phi \psi \psi} $ or $g_{\varpi \psi \psi} $     and dimensionful coupling coefficients 
      $M_{\phi gg} $ or $M_{\varpi gg} $
     to the gluon scalar or pseudoscalar kinetic function normalized to the Higgs field 
       expectation value.   
       The branching ratios in rows four and eight assume $M_{\phi gg}, M_{\varpi gg} \gsim v$. 
              }
\label{table:S} 
\end{center}
\end{table}


One of the simplest limiting patterns of marginal di-fermion couplings for a spin-zero boson 
  is to di-tau and di-top only. 
 This is illustrated in the first and fifth rows of Table 1 for equal $g_{\phi \tau \tau } = g_{\phi tt}$ 
  or $g_{\varpi \tau \tau } = g_{\varpi tt}$. 
 In this case the di-tau branching is nearly unity, with only a small correction from di-gluon 
  parton decays through a top quark loop.  
  Direct production in proton-proton collisions is from gluon fusion through a top quark loop. 
  With the upper limit restrictions of roughly unity on the $g_{\phi tt}$ or $g_{\varpi tt}$ 
  di-quark dimensionless coupling coefficients  
   specified above, the total production cross section for either scalar $\phi$ or
    pseudoscalar $\varpi$ is limited in this case to the nanobarn range, 
   and the cross section times branching ratio to di-tau final state 
   into the benchmark kinematic fiducial region is
   limited to roughly the one-tenth 
  nanobarn level. 
  
 Another simple limiting pattern of di-fermion couplings of spin-zero bosons is to di-tau and di-bottom only. 
  This is illustrated in the second and sixth rows of Table 1 for equal $g_{\phi \tau \tau } = g_{\phi bb}$ 
  or $g_{\varpi \tau \tau } = g_{\varpi bb}$. 
 In this case the di-tau branching ratio is also nearly unity. 
 Direct production in proton-proton 
  collisions is dominantly by gluon fusion through a bottom quark loop, 
   with additional production in association with a bottom quark and anti-bottom quark 
    down by roughly an order of magnitude.  
  The total direct production cross section with only di-bottom coupling 
    is similar to that of the case above with only di-top coupling 
for similar values 
     of the dimensionless coupling coefficients $g_{\phi bb}$ and $g_{\phi tt}$ or 
    $g_{\varpi bb}$ and $g_{\varpi tt}$.   
    This is because for fixed dimensionless coupling coefficient, 
    the integrated gluon fusion scattering probability to a spin-zero 
   boson becomes roughly independent of loop quark mass for masses of order or larger 
    than the boson mass, which is roughly applicable here  
    for both bottom and top quark loops with  $m_{\phi, \varpi} = 10$ GeV.    
 However, contributions to the transverse momentum spectrum of the boson 
 coming from gluons emitted from the quark loop 
  are significantly softer in the region $p_T^{\phi , \varpi} \gsim m_\phi$ 
 for the lighter bottom quark than the much heavier top quark.  
 For $m_{\phi, \varpi} = 10$ GeV this results in an acceptance within the benchmark kinematic fiducial region 
 of $p_{T}^{\phi , \varpi } \! > \! 20 \, {\rm GeV} $ nearly 
    an order of magnitude smaller for a bottom quark loop compared to the case with a top quark loop. 
But with the
 upper limit restrictions of roughly $m_t / m_b$ on the $g_{\phi bb}$ or $g_{\varpi bb}$ 
  di-quark dimensionless coupling coefficients  
   specified above, the  cross section times branching ratio to di-tau final state 
   into the benchmark kinematic fiducial region 
   for either scalar $\phi$ and 
    pseudoscalar $\varpi$ can reach the many tens of nanobarns level 
    in this case. 
Relaxing the restrictions specified above on the di-quark dimensionless coupling coefficients 
    for first and second generation quarks can 
    yield similar or even larger levels. 
    
The final  
simple limiting pattern of marginal di-fermion couplings for a spin-zero scalar or pseudoscalar boson 
presented in rows three and seven in Table 1 is for equal dimensionless coupling coefficients 
$g_{\phi \psi \psi }$ or $g_{\varpi \psi \psi}$
to all di-fermions. 
This pattern realizes universal exact 
mass-proportional chirality-violating di-fermion 
interactions with a spin-zero boson across all fermion types. 
 In this case the boson branching ratios to di-tau and parton di-charm are roughly equal, with  
 only a small correction from parton di-gluon.  
 The total direct production cross section in proton-proton collisions is roughly a factor of four 
  larger than the case with top quark coupling above, due to net constructive interference 
between top and bottom quark loops 
     in gluon fusion production. 
 With the upper limit restrictions specified above applicable 
  to this case of roughly unity on the $g_{\phi \psi \psi}$ or $g_{\varpi \psi \psi}$ 
  di-quark dimensionless coupling coefficients,  
   the cross section times branching ratio to di-tau final state 
   into the benchmark kinematic fiducial region for either scalar $\phi$ or 
    pseudoscalar $\varpi$ is
   limited to roughly the one-tenth of 
    a nanobarn level. 
      For reference, the di-muon and di-photon branching ratios with this pattern of equal 
      dimensionless coupling coefficients for all fermions are 
 $ {\rm Br}( \phi , \varpi  \to \mu \mu ) \simeq (2.4 \, , \, 2.0)  \times10^{-3}$
  and 
 $ {\rm Br}( \phi , \varpi  \to \gamma \gamma) \simeq ( 3.0 \, , 4.7 )  \times10^{-5}$ 
 respectively.

Another class of models  with a spin-zero scalar or pseudoscalar boson
that can contribute di-lepton non-universality in the bottomonium mass region 
 are ones with marginal chirality-violating 
 di-fermion couplings (\ref{scalarcouplings}) or (\ref{pseudoscalarcouplings}) only to di-leptons, 
 without any direct di-quark couplings. 
 In this case production of a spin-zero boson through gluon fusion 
  in proton-proton collisions can come from 
  interactions that are  quadratic in the gluon field  
  strength. 
  The leading most relevant  allowed 
  interaction of this type with a single 
   real spin-zero scalar boson $\phi$ is proportional to the scalar gluon kinetic density
\beq
 { \alpha_s \over 12 \pi   }    {\phi \over M_{\phi gg}  } 
  ~G^a_{\mu \nu} G^{a \mu \nu}
  \label{scalarGG}
\eq
where 
$G^a_{\mu \nu}$ is the gluon field strength, 
$M_{\phi gg} $ is an ultraviolet mass scale, and 
the one-loop prefactor is chosen for convenience. 
 For reference,  in the case where the scalar boson is the Higgs boson, $\phi = h$,  
 this interaction is obtained with $M_{\phi gg} =v$ for a single quark flavor that gains 
  mass from electroweak symmetry breaking in the heavy quark limit. 
 Likewise,  the leading most relevant  allowed 
 interaction of this type with a single 
   real spin-zero pseudoscalar boson $\varpi$, 
   again often referred to  
 as an axion-like particle,
    is proportional to the pseudoscalar gluon kinetic density
\beq
  { \alpha_s \over 8 \pi} {\varpi \over M_{\varpi gg}  }
  ~G^a_{\mu \nu} \tilde{G}^{a \mu \nu}
  \label{pseudoscalarGG}
  \eq
where 
$\tilde{G}^a_{\mu \nu} = {1 \over 2} \epsilon_{\mu \nu \rho \sigma} G^{a \rho \sigma }$ is the gluon 
 dual field strength,  
$M_{\varpi gg} $ is an ultraviolet mass scale, and 
the one-loop prefactor is chosen for convenience in matching to an 
 ultraviolet completion described below. 
 For both interactions above, by definition $\phi=0$ or $\varpi =0$ in the vacuum. 
 
The branching ratios for a spin-zero scalar or pseudoscalar boson  
with marginal di-lepton couplings  (\ref{scalarcouplings}) or (\ref{pseudoscalarcouplings})
and irrelevant couplings quadratic in gluon field strength
(\ref{scalarGG}) or (\ref{pseudoscalarGG}) 
 depend on the di-lepton 
  dimensionless coupling coefficients $g_{\phi \ell \ell }$ or $g_{\varpi \ell \ell}$
 and  the ultraviolet scale $M_{\phi gg} $ or $M_{\varpi gg} $. 
 With approximately mass-proportional di-lepton couplings, 
  with dimensionless coupling coefficients not differing significantly from unity   
 among the three lepton flavors,  and for a gluon interaction ultraviolet mass scale 
  near or above the electroweak scale, 
  the di-tau final state dominates the boson branching ratios,  
 with only small corrections from di-gluon parton decays. 
 Spin-zero boson production in proton-proton collisions in this case without direct coupling 
  to di-quarks of course comes only from 
  gluon fusion. 
 The results in this class of models for di-tau branching ratio  
 ${\rm Br}( \phi, \varpi \to \tau \tau)$,  
 the total inclusive cross section 
 $\sigma(pp \to \phi, \varpi) $,
  and cross section times branching ratio times acceptance in the benchmark fiducial kinematic region 
 $\sigma(pp \to \phi , \varpi) \cdot  {\rm Br}( \phi, \varpi \to \tau \tau) \, \cdot 
    {\rm Acc}(p_{T}^{\phi, \varpi } \! > \! 20 \, {\rm GeV} \, , \, | y_{\phi, \varpi} | < 1.2) $, 
  are presented 
  in rows four and eight 
in Table 1  for scalar or pseudoscalar bosons of mass 10 GeV.   
 The cross section times branching ratio to di-tau final state 
   into the benchmark kinematic fiducial region is
   limited to roughly the one-tenth nanobarn level 
   for $M_{\phi gg} $ or $M_{\varpi gg} $ near the electroweak scale, 
    and to roughly the one-hundredth nanobarn level at the TeV scale.



\subsection{Spin-Zero Boson Ultraviolet Completions}

There are a number of 
 classes 
 ultraviolet completions 
for the underlying 
origin of the 
interactions of a spin-zero scalar or pseudoscalar 
 boson with SM fermions and massive gauge bosons in 
(\ref{scalarcouplings}) or (\ref{pseudoscalarcouplings}). 
One is that there are no direct couplings in the ultraviolet 
 of the spin-zero boson to SM fermions or gauge bosons, 
  but rather only 
  couplings through the SM (extended) Higgs sector.  
  In this case interactions with SM fermions and gauge 
  bosons are inherited from SM Yukawa and gauge interactions 
   through mixing of the spin-zero boson with (extended) Higgs sector spin-zero states.   
To linear order in the spin-zero boson field, all such interactions 
 are proportional to the interactions of the 
 Higgs sector states. 
 The constant of proportionality is characterized at leading order entirely 
  in terms of a mixing amplitude $\sin \theta$ of the spin-zero boson with the appropriate Higgs sector state(s). 
The mixing amplitudes squared in all cases are limited by  $\sin^2 \theta \leq {1 \over 2}$ 
 where the inequality is saturated for maximal mixing.

The simplest Higgs sector is the single Higgs Doublet Model (HDM) of the SM 
with the spin-zero Higgs boson $h$. 
With an additional spin-zero boson 
a Higgs Doublet Mixing Model (HDMM) can result from 
Higgs sector interactions  that induce 
mixing between the SM Higgs boson and the additional boson.  
With time-reversal symmetry only a spin-zero scalar boson $\phi$ can mix 
  with $h$. 
  In this case all of the fermion 
 dimensionless coupling coefficients in (\ref{scalarcouplings}) 
 are equal at leading order to the single mixing amplitude with the Higgs boson   
 $ g_{\phi \psi \psi}  = \sin \theta_{ \phi h}$.
 This sub-class of ultraviolet completion 
  realizes universal exact 
mass-proportional chirality-violating di-fermion 
interactions with a spin-zero scalar boson across all fermion types. 
It therefore represents an ultraviolet completion of the pattern of di-fermion 
 couplings given in the third row of Table 1. 
  In addition, it has associated mass squared proportional couplings 
  to massive SM gauge bosons in (\ref{scalarcouplings})   
  with the same mixing amplitude dimensionless coupling coefficients 
  $g_{\phi ZZ} = g_{\phi WW} =  \sin \theta_{ \phi h}$. 
   The coupling of the spin-zero scalar boson to massive SM gauge bosons 
  in this sub-class introduces other signals for which  
   there are experimental constraints.  
 In particular, the existence of a light spin-zero scalar boson 
  with decays to di-tau
 produced in association with a $Z$ boson is bounded from searches
  at the LEP II electron-positron collider \cite{LEPWorkingGroupforHiggsbosonsearches:2003ing}. 
 For $m_\phi = 10$ GeV the bound  in 
  terms of the 
  $Z$ boson dimensionless coupling coefficient is  
   $g^2_{\phi ZZ} \cdot {\rm Br}(\phi \to \tau \tau) < 0.09 $ 
 at 
 the 95\% confidence level \cite{LEPWorkingGroupforHiggsbosonsearches:2003ing}.  
 In the case here 
  with the di-tau branching ratio of ${\rm Br}(\phi \to \tau \tau) \simeq 0.56$ 
 from Table 1, this amounts to a bound on the mixing amplitude 
  squared of $\sin^2 \! \theta_{\phi h} < 0.16$.  
  With this bound, the direct production cross section times branching ratio  
  to di-tau final state 
   into the benchmark kinematic fiducial region given in the third row of Table 1
  in this sub-class of ultraviolet completions 
   is limited to roughly the one-hundredth nanobarn level. 
 
The simplest extended Higgs sector is the Two Higgs Doublet Model (2HDM) 
with three spin-zero bosons $h, H, A$. 
With an additional spin-zero boson 
a Two Higgs Doublet Mixing Model (2HDMM) can result from 
Higgs sector interactions  that induce 
mixings between the 2HDM boson states and the additional boson. 
 With time-reversal symmetry a spin-zero scalar boson $\phi$ can mix 
  with $h$ and $H$, or a spin-zero pseudoscalar $\varpi$ can mix with $A$.  
 This Two Higgs Doublet Mixing Model (2HDMM) sub-class of ultraviolet completions 
  is often employed to characterize the interactions of an additional  light 
  scalar \cite{Drozd:2014yla} 
  or pseudo-scalar \cite{Argyropoulos:2022ezr}
  spin-zero boson with SM fields.  
     
  The class of 2HDMMs considered here will be restricted to ones without 
   tree-level quark or lepton flavor changing interactions with neutral Higgs sector states. 
 This restriction is enforced by the Glashow-Weinberg condition \cite{Glashow:1976nt}
  that all SM fermions of a given representation under the SM gauge group, 
   namely up-type quarks, down-type quarks, and leptons, 
 receive their mass through marginal Yukawa couplings to a single Higgs doublet 
 \beq 
 - \lambda_U \, \bar{\psi}_U \, H_2^c \, P_R \, \psi_U 
 - \lambda_D \, \bar{\psi}_D \, H_i \, P_R \, \psi_D 
 - \lambda_\ell \, \bar{\psi}_\ell \, H_j \, P_R \, \psi_\ell 
  ~ + ~ {\rm h.c.} 
   \eq 
where 
$P_{L,R} = {1 \over 2} ( 1 \mp \gamma_5)$ and $H^c = i \sigma^2 H^*$,  
and the indices $i,j = 1$ or $2$ label the two Higgs doublet fields.  
 In this case tree-level couplings of all neutral Higgs sector states are diagonal 
  in the fermion mass eigenbases.  
 There are four discrete types of 2HDMMs that satisfy the Glashow-Weinberg condition.  
In types I and II 
 the Higgs doublet couplings to down-type quarks and leptons are $ i=j=2$ and 
 $i=j=1$ respectively, 
  while for types III and IV the doublet couplings are 
  $i=2, j=1$ and $i=1, j=2$ respectively. 
 By convention in all types, couplings to the up-type quarks 
 are with the second Higgs doublet field.


\begin{table}[t]  
\begin{center}  \vspace*{0.2in} 
\begin{tabular}{crrrr} 
  Dimensionless  & Type I   & Type II & Type III & Type IV \\
Coupling Coefficient  & &  &     \\
    & \\
 $g_{\phi UU}  /  \sin \theta_{\phi H}$  &  $- \cot \beta$   & $- \cot \beta$   & $- \cot \beta$ & $- \cot \beta$     \\
    & \\ 
 $g_{\phi DD}  /  \sin \theta_{\phi H}$  &  $- \cot \beta$   & $ \tan \beta$   & $- \cot \beta$ & $ \tan \beta$     \\
    & \\
 $g_{\phi \ell \ell}  /  \sin \theta_{\phi H}$  &  $- \cot \beta$   & $ \tan \beta$   & $ \tan \beta$ & $ - \cot \beta$     \\
    & \\
 & \\
  $g_{\varpi UU}  /  \sin \theta_{\varpi A}$  &  $ \cot \beta$   & $ \cot \beta$   & $ \cot \beta$ & $ \cot \beta$     \\
    & \\ 
 $g_{\varpi DD}  /  \sin \theta_{\varpi A}$  &  $- \cot \beta$   & $ \tan \beta$   & $- \cot \beta$ & $ \tan \beta$     \\
    & \\
 $g_{\varpi \ell \ell}  /  \sin \theta_{\varpi A}$  &  $- \cot \beta$   & $ \tan \beta$   & $ \tan \beta$ & $ - \cot \beta$     \\ 
 & \\
  \end{tabular}
\caption{  Two Higgs Doublet Mixing Model dimensionless   coupling coefficients divided by mixing amplitudes 
  for scalar $\phi$ or pseudoscalar $\varpi$ spin-zero bosons 
   to up-type quarks, down-type quarks, and leptons. 
  Scalar coupling coefficients are in the alignment limit.  
              }
\label{table:2HDMMcouplings} 
\end{center}
\end{table}

With the restrictions specified above, 
the couplings of the two Higgs doublet states $h, H, A$ to SM fermions and massive 
 gauge bosons are specified entirely 
 in terms of $\tan \beta = v_2 / v_1$ where $v_1$ and $v_2$ are the expectation values of the 
  two Higgs doublet fields, and  
  the overlap $\sin ( \beta - \alpha)$ of the SM-like Higgs mass eigenstate $h$ with the 
   expectation value eigenvector in the Higgs doublet space,  
    or equivalently 
     the overlap $\cos \alpha$ of this mass eigenstate with the second 
   Higgs doublet
     \cite{Gunion:2002zf,Craig:2013hca}.
The dimensionless coupling coefficients of a scalar $\phi$  or pseudoscalar $\varpi$
 spin-zero boson   
 to SM fermions and massive gauge bosons 
 in (\ref{scalarcouplings}) or (\ref{pseudoscalarcouplings})
 that 
 arise from mixing with these two Higgs doublet states 
 are then in turn specified in terms of $\tan \beta$ and $\sin(\beta- \alpha)$ along with the mixing 
  amplitudes $\sin \theta_{\phi h}$, $\sin \theta_{\phi H}$ or $\sin \theta_{\varpi A}$. 
 Since 
 up-type quarks, down-type quarks, and leptons, each
   couple to only a single Higgs doublet field, 
  there are three separate dimensionless coupling coefficients in (\ref{scalarcouplings}) or (\ref{pseudoscalarcouplings})
  for each of these types of fermions. 
    And each of these three dimensionless coefficients is independent of flavor.  
The 2HDMM sub-class of ultraviolet completions therefore realize mass-proportional 
  chirality-violating di-fermion interactions with a spin-zero scalar or pseudoscalar boson 
 with independent dimensionless constants of proportionality for each of the three types of massive SM Dirac fermions.  
  
  The dimensionless coupling coefficients in (\ref{pseudoscalarcouplings})
   for up-type quarks, down-type quarks, and leptons 
  for a spin-zero pseudoscalar boson $\varpi$ 
   in 
  any of the four types of 2HDMMs are simply equal to $\tan \beta$ or $\cot \beta$ up to a sign, times
  the mixing amplitude $\sin \theta_{\varpi A}$, as 
given in Table 2.  
In order to simplify the presentation of the dimensionless coupling coefficients  in (\ref{scalarcouplings}) 
for a scalar spin-zero boson $\phi$ 
it is useful to define a 2HDMM alignment limit in which the 
SM-like Higgs mass eigenstate is aligned with the expectation value eigenvector in the two Higgs 
 doublet space. 
This is analogous to the alignment limit in 
 2HDMs \cite{Craig:2013hca}. 
In this limit $\sin(\beta - \alpha)=1$ and $\cos(\beta - \alpha) =0$, and the scalar spin-zero mixing amplitude 
 with the SM-like Higgs boson vanishes, $\sin \theta_{\phi h} = 0$. 
The leading interactions of the SM-like Higgs $h$ with SM fermions and massive gauge 
 bosons are equal to those of the single Higgs doublet SM Higgs boson in this limit. 
 In addition, the dimensionless coupling coefficient of the scalar spin-zero boson 
  to 
  massive gauge bosons in (\ref{scalarcouplings}) vanishes, 
   $g_{\phi ZZ} = g_{\phi WW} = 0$. 
Away from this limit these  coupling are
$ g_{\phi ZZ} = g_{\phi WW} =  \sin \theta_{\phi h} \sin( \beta - \alpha) + \sin \theta_{\phi H} \cos(\beta - \alpha)$.  
The LEP II bound discussed above requires this dimensionless coupling to be somewhat small, 
 so it is reasonable to work in the 2HDMM alignment limit as a first approximation.  
In this 2HDMM alignment limit 
the dimensionless coupling coefficients in (\ref{scalarcouplings})
   for up-type quarks, down-type quarks, and leptons 
  for a spin-zero scalar boson $\phi$ 
   in 
  any of the four types of 2HDMMs are also equal to $\tan \beta$ or $\cot \beta$ up to a sign, times
  the mixing amplitude $\sin \theta_{\phi H}$, as 
given in Table 2.  

The restrictions discussed earlier ensuring 
that Yukawa-type couplings of 
third generation top and bottom quarks do not parametrically exceed their infrared quasi-fixed point values 
 applies in the 2HDMM sub-class of ultraviolet completions directly to the Yukawa couplings of third generation quarks 
  to the Higgs doublet fields. 
 In all four 2HDMM types this restricts $\tan \beta$ to be larger than roughly unity, and in type II and IV
 in addition restricts $\tan \beta$ to be smaller than roughly $m_t / m_b$.


\begin{figure}[htb] 
   \centering
   \includegraphics[width=5.5in]{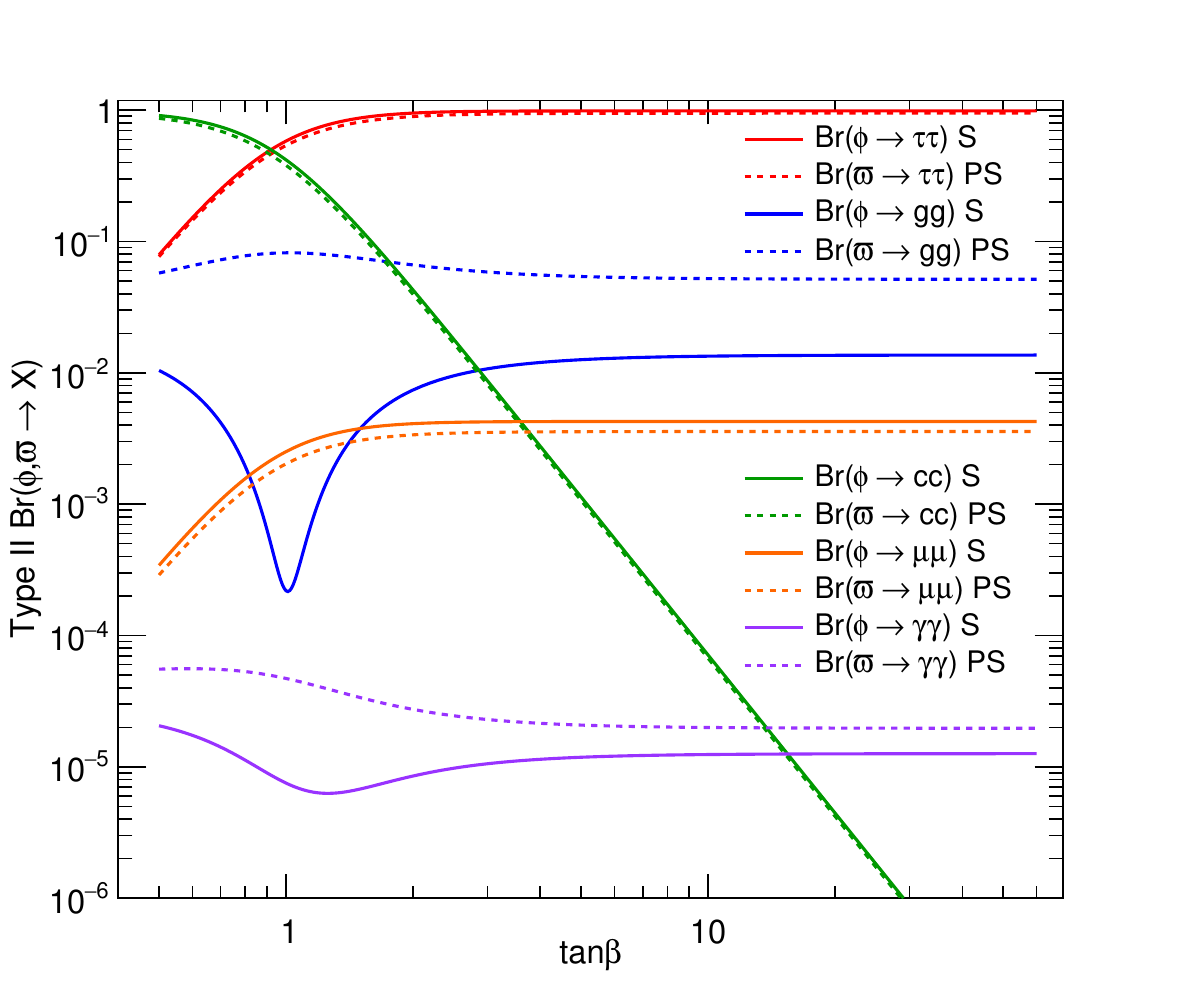} 
   \vspace*{-0.1in}
   \caption{Branching ratios of a spin-zero scalar $\phi$ or pseudoscalar $\varpi$ 
    boson of mass $m_{\phi, \varpi} = 10$ GeV 
     in type II Two-Higgs Doublet Mixing Models in the alignment limit 
     as a function of $\tan \beta$.             }
   \vspace{0.15in}
   \label{fig:1}
\end{figure}

The decay branching ratios of a spin-zero scalar or pseudoscalar boson  
in the alignment limit of the 
2HDMM sub-class of ultraviolet completions
is independent of mixing amplitudes since all partial decay widths are proportional 
 to the same mixing amplitude squared. 
 In all cases the dominant branching ratios are very similar for scalar $\phi$ or pseudoscalar $\varpi$. 
In type I the branching ratios are independent of $\tan \beta$ since all partial decay 
 widths are proportional to $\cot^2 \! \beta$. 
The di-tau branching ratio in this case for 
 $m_{\phi , \varpi} = 10$ GeV is roughly a half. 
In types II and III for $m_{\phi , \varpi} = 10$ GeV 
the di-tau decay mode is dominant with a branching ratio that approaches unity for $\tan \beta$ 
 larger than a few.  
 The type II branching ratios  ${\rm Br}( \phi, \varpi \to \tau \tau , cc, gg, \mu \mu , \gamma \gamma)$ 
 for $m_{\phi , \varpi} = 10$ GeV in the alignment limit
 are shown as a function of $\tan \beta$ 
 in Figure 1. 
 The dip in the parton di-gluon branching ratio near $\tan \beta$ of unity is from destructive 
  interference between the real dispersive parts of the top and bottom quark loops. 
 In type IV for $m_{\phi , \varpi} = 10$ GeV 
 the bottom quark loop contribution to the parton di-gluon decay mode dominates for $\tan \beta$ larger than a few, with the 
 di-tau branching ratio falling like $\cot^4 \! \beta$ for large $\tan \beta$.


\begin{figure}[thb] 
   \centering
   \includegraphics[width=5.5in]{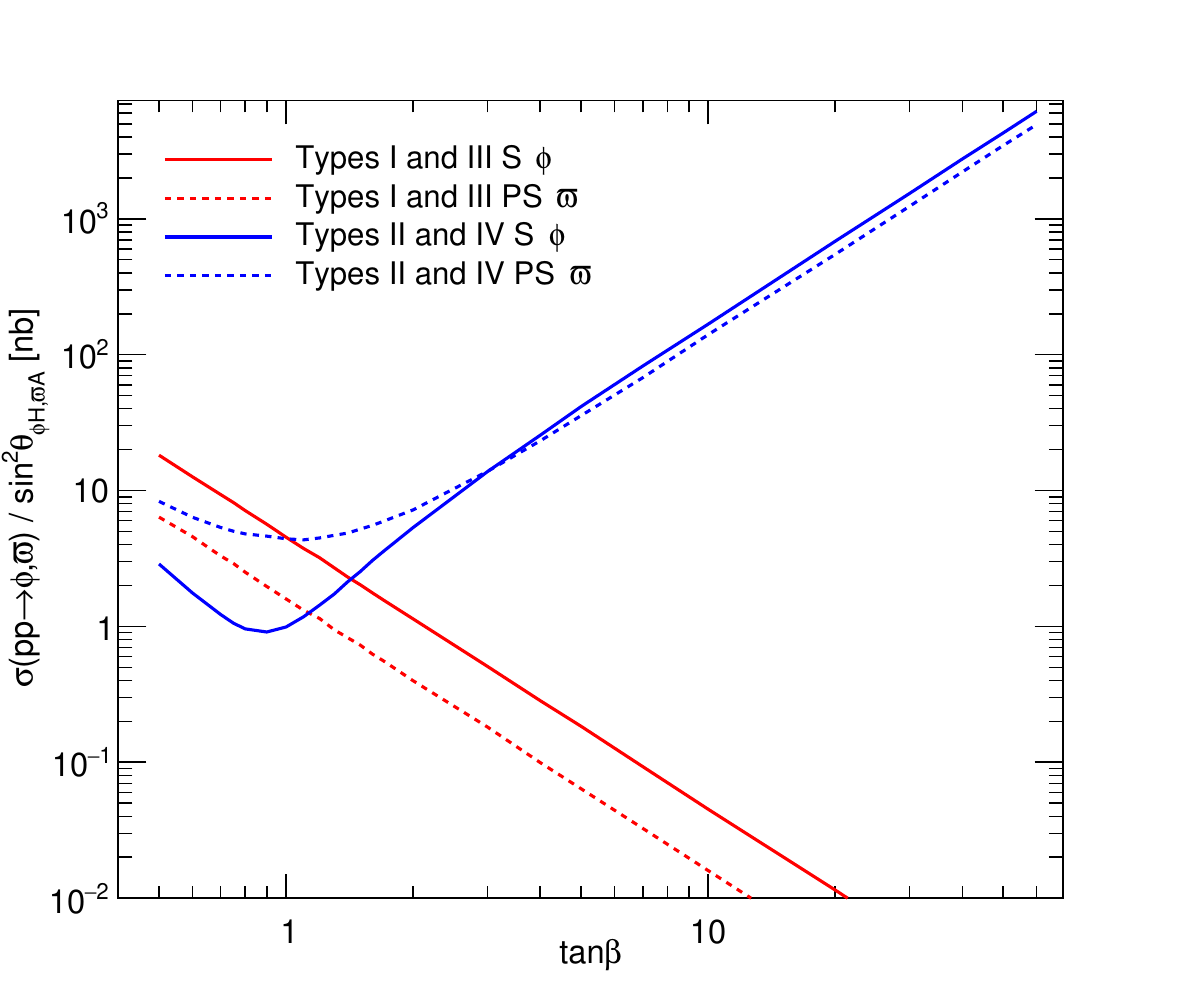} 
   \vspace*{-0.1in}
   \caption{Total inclusive production cross section 
   divided by mixing amplitude squared 
   $\sigma(pp \to \phi , \varpi ) / \sin^2 \theta_{\phi H , \varpi A}$
   in nanobarns 
   for the combined processes of gluon fusion and production in association 
     with a bottom and anti-bottom quark 
      of a spin-zero scalar $\phi$ or pseudoscalar $\varpi$ 
    boson of mass $m_{\phi, \varpi} = 10$ GeV 
     in 13.6 TeV proton-proton collisions 
     for Two-Higgs Doublet Mixing Models in the alignment limit 
     as a function of $\tan \beta$.  
         }
   \vspace{0.15in}
   \label{fig:2}
\end{figure}

Direct
production in proton-proton collisions for a scalar $\phi$ or pseudoscalar $\varpi$ 
in the 2DHMM sub-class of ultraviolet completions 
 is dominantly by gluon fusion through a top and bottom quark loop, 
   with additional direct production in association with a bottom quark and anti-bottom quark 
    down by roughly an order of magnitude.  
The total production cross sections divided by mixing amplitude squared 
 $\sigma(pp \to \phi, \varpi) / \sin^2 \theta_{\phi H, \varpi A}$,
for all four 2HDMM types in the alignment limit are shown for $m_{\phi , \varpi} = 10$ GeV
as a function of $\tan \beta$ in Figure 2. 
In types I and III 
the scalar production cross sections are identical, 
 as are the pseudoscalar cross sections, 
  all proportional to $\cot^2 \beta$. 
 Likewise 
in types II and IV the scalar total cross sections are identical, 
as are the pseudoscalar cross sections.
  The dip in the scalar cross sections 
  in these cases is in gluon fusion from destructive interference  
   between the real dispersive parts of the top and bottom quark loops. 
For $\tan \beta$ larger than a few in types II and IV 
the dominant direct production comes from coupling 
  to the bottom quark with total cross section scaling like $\tan^2 \beta$, 
  and the top quark loop contribution to gluon fusion falling like $\cot^2 \beta$.


\begin{figure}[th] 
   \centering
   \includegraphics[width=5.5in]{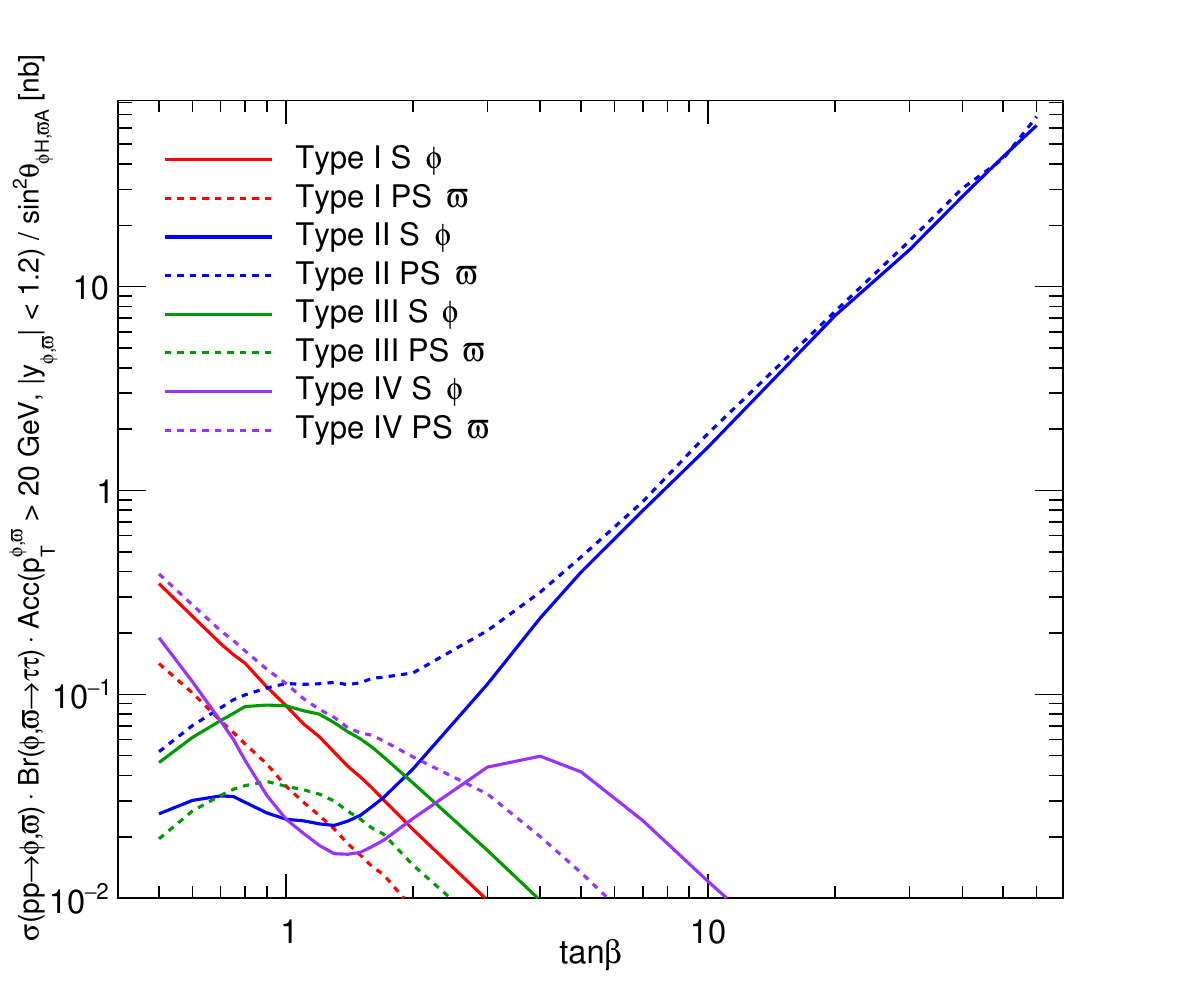} 
   \vspace*{-0.1in}
   \caption{
        Total inclusive production cross section 
        times di-tau branching ratio times acceptance divided by mixing amplitude squared  
   $\sigma(pp \to \phi , \varpi ) \cdot {\rm Br}( \phi, \varpi \to \tau \tau) 
    \cdot {\rm Acc}(p_{T}^{\phi , \varpi} \! > \! 20 \, {\rm GeV} \, , \, | y_{\phi, \varpi} | < 1.2) 
   /  \sin^2 \theta_{\phi H , \varpi A}  $
   in nanobarns 
   for the combined processes of gluon fusion and production in association 
     with a bottom and anti-bottom quark 
      of a spin-zero scalar $\phi$ or pseudoscalar $\varpi$ 
    boson of mass $m_{\phi, \varpi} = 10$ GeV 
     in 13.6 TeV proton-proton collisions 
     for Two-Higgs Doublet Mixing Models in the alignment limit 
     as a function of $\tan \beta$.  
     }
   \vspace{0.15in}
   \label{fig:3}
\end{figure}

The spin-zero scalar $\phi$ and pseudoscalar $\varpi$
direct production cross section times branching ratio times acceptance in the benchmark fiducial kinematic region 
divided by the mixing amplitude squared 
     $\sigma(pp \to \phi , \varpi) \cdot  {\rm Br}( \phi, \varpi \to \tau \tau) \, \cdot 
    {\rm Acc}(p_{T}^{\phi, \varpi } \! > \! 20 \, {\rm GeV} \, , \, | y_{\phi, \varpi} | < 1.2) / \sin^2 \theta_{\phi H, \varpi A}$, 
 for all four 2HDMM types in the alignment limit are shown for 
    $m_{\phi , \varpi} = 10$ GeV
as a function of $\tan \beta$ in Figure 3. 
In all four types for $\tan \beta$ of order unity with near maximal mixing amplitude,
the direct production cross section times branching ratio  
  to di-tau final state 
   into the benchmark kinematic fiducial region for either scalar or pseudoscalar spin-zero boson 
     is limited to roughly the one-tenth to one-hundredth nanobarn level. 
Only for type II 2HDMMs do the scalar and pseudoscalar direct production cross sections
  times di-tau branching ratio grow like $\tan^2 \beta$ 
 for $\tan \beta$ larger than a few. 
  In this case the dominant scalar or pseudoscalar di-fermion couplings are to di-tau and di-bottom.  
 In the large $\tan \beta$ limit,  
 type II 2HDMMs 
 asymptote to the pattern of 
 dimensionless coupling coefficients given in the second and sixth rows of Table 1 
  with $g_{\phi \tau \tau} = g_{\phi bb} = \sin \theta_{\phi H} \tan \beta$ 
   or $g_{\varpi \tau \tau} = g_{\varpi bb} = \sin \theta_{\varpi A} \tan \beta$. 
For moderately large mixing amplitudes and large enough $\tan \beta$, 
 the direct production cross section times di-tau branching ratio times acceptance in the benchmark 
 fiducial kinematic region  can reach the many nanobarn level 
  for either scalar $\phi$ or pseudoscalar $\varpi$ 
spin-zero bosons in the type II 2HDMM sub-class of ultraviolet completions. 

Another class of 
ultraviolet completions 
of the interactions of a spin-zero scalar $\phi$ or pseudoscalar $\varpi$ 
 boson with SM fermions  in 
(\ref{scalarcouplings}) or (\ref{pseudoscalarcouplings}) 
is directly from irrelevant Yukawa couplings. 
While SM gauge symmetry does not allow marginal Yukawa couplings of SM 
 fermions with a gauge singlet spin-zero boson in the ultraviolet, such couplings can arise from 
 allowed irrelevant Yukawa couplings with an additional spin-zero field $\Phi$ 
  of the form 
\beq 
 - {\lambda_{\Phi \psi} \over M} \, \Phi \, \bar{\psi} \, H \, P_R \, \psi 
  ~ + ~ {\rm h.c.} 
  \label{irryuk}
\eq 
where $H = H^c$ for up-type quarks. 
In the vacuum with a non-zero expectation value $f = \sqrt{2} \, \langle \Phi \rangle $ 
these irrelevant Yukawa couplings give an effective Yukawa coupling 
of SM fermions to the Higgs doublet field with 
 $\lambda_\psi = \lambda_{\Phi \psi}  f / \sqrt{2} \, M$. 
 In the vacuum with massive electroweak gauge bosons with Higgs doublet expectation value, 
  spin-zero excitations of $\Phi$ then inherit effective Yukawa couplings to SM fermions 
 that have  irrelevant Yukawa couplings of the form (\ref{irryuk}). 
As long as each type of SM fermion gets a mass either from a standard 
marginal Yukawa coupling with a single Higgs 
 doublet field, or from an irrelevant Yukawa coupling (\ref{irryuk}) with a single Higgs doublet field, 
 but not both types, then 
 tree-level couplings of all $\Phi$ excitations and neutral Higgs sector states are diagonal
in the fermion mass eigenbases.
This condition is analogous to the Glashow-Weinberg condition discussed above for 2HDMs or 2HDMMs. 
Alternately, if each individual fermion gets a mass from either a standard marginal Yukawa coupling or from 
  an irrelevant Yukawa coupling (\ref{irryuk}) but not both, possibly enforced by (discrete) flavor symmetries,
  then all Yukawa interactions commute and the same result is achieved.

The interactions of a light spin-zero 
 boson with SM fermions  in 
(\ref{scalarcouplings}) or (\ref{pseudoscalarcouplings}) 
can arise from irrelevant Yukawa couplings (\ref{irryuk}) either through mixing 
 with, or directly as, an excitation of the $\Phi$ field. 
Mixing of a spin-zero scalar boson $\phi$ 
 with the Higgs mode $\eta$, where $ \sqrt{2} \, \Phi = f + \eta$, with mixing amplitude 
$\sin \theta_{\phi \eta}$ gives an interaction with SM fermions that gain a mass from the 
 irrelevant Yukawa coupling (\ref{irryuk}) with dimensionless coupling coefficient in (\ref{scalarcouplings}) 
 of  $g_{\phi \psi \psi} =  \sin \theta_{\phi \eta}  \, v / f$. 
In the case that $\Phi$ is a complex field, the angular mode $\varpi$, 
 where  $ \sqrt{2} \, \Phi = f \, e^{i \varpi / f}$,
is a spin-zero pseudoscalar boson with 
 inherited effective coupling to SM fermions with dimensionless coupling coefficient in (\ref{pseudoscalarcouplings}) 
of $g_{\varpi \psi \psi} =   v / f$. 
In this case the pseudoscalar boson $\varpi$ could be a light pseudo-Goldstone boson 
 as the result of an 
 approximate spontaneously broken Abelian global symmetry in the $\Phi$ sector.  
 It is worth noting that the $\Phi$ expectation value $f$ could in principle be smaller 
  than the electroweak Higgs doublet expectation value  $v$. 
For light SM fermions including the leptons and bottom quark, the dimensionless coupling coefficients
to di-fermions that gain a mass from the irrelevant Yukawa couplings (\ref{irryuk})  could then exceed unity. 
This is particularly true in the case that $\varpi$ is a light pseudo-Goldstone boson 
 without any additional mixing amplitude suppression in the dimensionless coupling coefficients. 


Models with massive vector-like fermions that mix with SM fermions represent a sub-class of ultraviolet completions 
 that can give rise to irrelevant Yukawa couplings of the form (\ref{irryuk}). 
In this Vector-Like Fermion Mixing Model (VLFMM) sub-class 
a massive vector-like Dirac fermion $\psi'$ 
that has the same SM gauge representation as the right-handed component of a 
 SM Dirac fermion $\psi$, 
has allowed marginal Yukawa couplings to the SM Dirac fermion both through a Higgs doublet field $H$
 as well as a SM gauge-singlet spin-zero field $\Phi$ 
 \beq
- 
              \lambda'_{H \psi}  \, \bar{\psi} \, H \, P_R \, \psi'  
            - \lambda'_{\Phi \psi} \,  \bar{\psi}' \, \Phi \, P_R \,   \psi ~ + {\rm h.c.} \, 
      \label{VLFyuk}
\eq
Integrating out the massive vector-like Dirac fermion with mass $m_{\psi'}$ generates the 
 irrelevant Yukawa coupling (\ref{irryuk}) with 
$ \lambda_{\Phi \psi} / M = \lambda'_{H \psi} \, \lambda'_{\Phi \psi}  / m_{\psi'}$. 
As long as each individual 
SM fermion gets a mass either from a standard marginal Yukawa coupling with a single Higgs 
 doublet field, or from 
 the combination of marginal Yukawa couplings (\ref{VLFyuk}) with massive vector-like fermions 
  and single Higgs doublet and $\Phi$ fields, 
 but not both types, then 
 tree-level couplings of all $\Phi$ excitations and neutral Higgs sector states  are diagonal
in the fermion mass eigenbases.
The  specific form of the allowed marginal Yukawa couplings (\ref{VLFyuk}) as well as disallowed 
 standard marginal Yukawa couplings with Higgs doublet fields, may be enforced 
by (discrete) symmetries.  
A massive vector-like Dirac fermion 
that has the same SM gauge representation as the left-handed component of a 
 SM Dirac fermion can also generate irrelevant Yukawa couplings of the form (\ref{irryuk}). 
 The specific pattern of dimensionless coupling coefficients for spin-zero bosons that either mix with, 
  or directly are, excitations of the $\Phi$ field depends on details of the ultraviolet completion in these
  VLFMMs.     
     
Massive Vector-Like Quark Models (VLQMs) represent a related sub-class of ultraviolet completions for 
 a spin-zero boson without direct SM di-quark interactions, 
  but which can still be produced  through gluon fusion through coupling to
  either the scalar (\ref{scalarGG}) or pseudoscalar (\ref{pseudoscalarGG}) gluon kinetic densities.   
In this sub-class 
$N_f$ flavors of vector-like Dirac quarks have marginal Yukawa couplings to a spin-zero field $\Phi$
\beq
- \lambda_{\Phi \psi_Q} \, \Phi \, \bar{\psi}_Q \, \psi_Q  
\eq
The vector-like quarks gain all of their mass $m_{\psi_Q} = \lambda_{\Phi \psi_Q} \, f / \sqrt{2}$
 from the expectation value of $\Phi$. 
Integrating out the massive vector-like quarks gives a one-loop coupling 
 of the Higgs mode excitation $\eta$ of the field $\Phi$ to the scalar gluon kinetic density.  
And mixing of  a spin-zero scalar boson $\phi$ 
 with this Higgs mode with mixing amplitude $\sin \theta_{\phi \eta}$ then gives 
 the interaction (\ref{scalarGG}) with irrelevant coupling 
  coefficient $1/M_{\phi gg} = \sin \theta_{\phi \eta} \, N_f / f$.  
 In the case that $\Phi$ is a complex field, integrating out the massive vector-like quarks gives a 
  one-loop coupling of 
  the angular pseudoscalar boson mode $\varpi$
 to the pseudoscalar gluon kinetic density (\ref{pseudoscalarGG})
 with irrelevant coupling coefficient $1/M_{\varpi gg} = N_f / f$.   
 As above, in this case 
  the pseudoscalar boson $\varpi$ could be a light pseudo-Goldstone boson 
 as the result of an 
 approximate spontaneously broken Abelian global symmetry in the $\Phi$ sector, 
  with no additional mixing amplitude suppression in the coupling to the pseudoscalar 
   gluon kinetic density.



\section{Conclusions} 

Searches at the LHC for signs of resonant new physics bosons in the 
 di-muon  spectrum have drastically reduced sensitivity 
 for masses near known SM composite di-muon resonances. 
The recent advent of observations and measurements of the di-electron and di-tau mass
spectra in the bottomonium mass region, made possible through the 
use of low-threshold high band-width triggers 
  with compressed event-level information, 
alone suffer the same limitations. 
In this paper we have pointed out that the inherent limitations in the search for new boson  
 resonances in  individual di-lepton 
measurements in the bottomonium mass region at the LHC
 may be overcome through simultaneous measurements of prompt di-lepton 
 spectra above the continuum background
  across more than one lepton flavor. 
SM contributions to 
prompt di-lepton production throughout the entire bottomonium mass region 
 is expected to be di-lepton flavor universal to a high degree. 
  A simultaneous measurement such as the one proposed here could be sensitive to 
new boson resonances with non-universal di-lepton decays
that lie within the bottomonium mass region. 
It could also be sensitive to non-SM decays of bottomonium states to non-universal di-lepton decay modes.  

Enhanced decay branching ratios to di-tau final states over di-muon and di-electron  
 arise naturally for new physics spin-zero bosons 
 with chirality-violating couplings to di-leptons. 
A measurement of the di-tau spectrum, with leptonic decay of one of the tau-leptons, 
has limited visible mass resolution and essentially 
 integrates over the entire bottomonium mass region.  
So tests for di-lepton non-universality in the integrated prompt di-lepton spectra above continuum background, 
 specifically for excess di-tau production, 
could be sensitive to either new spin-zero bosons 
anywhere in the bottomonium mass region, 
or to bottomonium states with non-SM enhanced di-tau decays.

Ideally, measurement and comparison of resonant prompt di-lepton mass spectra 
in the bottomonium mass region at the LHC
  would be carried out within a single analysis using a single trigger 
   and common di-lepton kinematics across flavor.  
Direct comparison of di-electron, di-muon, and di-tau channels utilizing existing individual measurements is 
 at present not possible as a practical matter due to inherent flavor dependence of kinematic acceptances. 
    In particular, the unique acceptance characteristics of the di-tau channel reconstructed through 
   its combined leptonic and hadronic decay modes do not match those imposed in di-electron and di-muon channels. 
   Direct tests of di-lepton universality in the bottomonium mass region will have to wait for future 
    dedicated measurements of the type proposed here with careful accounting for differences in kinematic 
     acceptance across lepton flavors.

 It is worth noting that although the target of the measurements proposed here 
  are prompt di-leptons arising from direct decays of new or known 
   resonances within the bottomonium mass region, there 
  are potential non-prompt backgrounds that also have a mass scale tied to the bottomonium region.  
   In the di-tau channel in particular, there are 
    decays of heavy bottomonium states to  
 a bottom and anti-bottom meson pair, where one meson decays leptonically 
  and the other decays either leptonically or hadronically. 
   Such decays can have similar topology to direct di-tau production when one tau is captured 
    through its muonic or electronic decay and the other is reconstructed hadronically. 
    And while these decays are technically non-prompt, since the bottom meson lifetimes are 
    only a few times longer than that of the tau-lepton, track displacement from a primary vertex 
    may not be a strong discriminator.  
    The decay chain kinematics are different in detail however, and could provide additional discrimination.   
 These non-prompt bottomonium backgrounds should be fully characterized and controlled in order to 
 deduce the di-tau contribution to prompt decays coming directly from resonances within the bottomonium mass 
  region. 

New physics resonant bosons that appear through signs of di-lepton non-universality in the bottomonium mass region 
 could have additional signals that might be probed at the LHC. 
   An experimentally interesting decay mode for a spin-zero boson in this mass region 
    is di-photon.
  The branching ratio depends on details of the ultraviolet completion, but for many of the universality 
   classes presented above $ {\rm Br}( \phi , \varpi  \to \gamma \gamma) \simeq {\rm few} \times10^{-5}$.   
  Searches for di-photon resonances above continuum background particularly in the bottomonium mass 
  region could face challenges from $\Upsilon(nS) \to ee$ decays with both the electron and positron faking photons.    
  
The degree of di-lepton non-universality that can arise at the LHC in the bottomonium mass
 region from a new physics boson with enhanced di-tau decays  depends on  
 details of the sub-class of ultraviolet completion. 
For the universality classes presented here,  the direct
 cross section times di-tau branching ratio times acceptance into the benchmark fiducial kinematic region 
  ranges from the one-hundredth of a nanobarn level for gluon fusion production of a spin-zero boson 
   coupled to di-leptons and   
  heavy vector-like quarks at the TeV scale, up to the many nanobarn level for a spin-zero boson with couplings 
   to SM quarks and leptons inherited from SM Yukawa couplings 
    through significant mixing with extended Higgs sector states with large 
    ratios of Higgs doublet expectation values. 
Tests of di-lepton universality in the bottomonium mass region 
 provide a compelling opportunity to uncover new physics that has, until now, remained hidden from view 
  at the LHC.


\vspace{0.4in} 

\noindent {\bf \Large Acknowledgements} 
\vspace{0.1in} 

\noindent
We would like to thank Matthew Buckley and David Shih for sharing their results 
 on pseudoscalar boson production in the bottomonium mass region.  
 This work was supported in part by the National Science Foundation under grant PHY 2512783, 
  and by the state of New Jersey through support for the 
   New High Energy Theory Center at Rutgers University.


\appendix


\vspace{0.5in} 

\section{Decay Widths} 

The decay width of a massive spin-zero boson $\varphi$ to a massive Dirac fermion--anti-fermion pair $\psi \bar{\psi}$
 with scalar couplings 
(\ref{scalarcouplings}) or pseudoscalar couplings (\ref{pseudoscalarcouplings}) 
 for $\varphi = \phi$ or $ \varpi$ is  
\beq
\Gamma( \varphi \to \psi \bar{\psi} ) 
 =  g_{\varphi \psi \psi}^2 \, N_\psi \, \,  {\bar{m}_\psi^2\,  m_{\varphi} \over 8 \pi \, v^2 }
\left( 1 - {4 m_\psi^2 \over m_{\varphi}^2 } \right)^{n_\varphi / \, 2}
\eq
where $m_\psi$ is the fermion pole mass for the kinematic phase space factor, 
$\bar{m}_\psi$ is the $\overline{\rm MS}$ mass at renormalization scale $\mu = m_\varphi$ 
for the coupling,
 and $N_\psi =1,3$ is the final state multiplicity for leptons, quarks. 
For pseudoscalar decay the massive fermions are $S$-wave near threshold with $n_\varpi = 1$, 
  while for scalar decay the fermions are $P$-wave near threshold with $n_\phi = 3$. 

The decay width of a massive spin-zero boson $\varphi$ to a gluon pair $gg$
arising at one-loop from a coherent sum over massive Dirac quarks $\psi_q$ with 
scalar couplings 
(\ref{scalarcouplings}) or pseudoscalar couplings (\ref{pseudoscalarcouplings})  
 for $\varphi = \phi$ or $ \varpi$ is  
 \beq
\Gamma( \varphi \to gg) =  C_\varphi \, {\alpha_s^2 ~m_\varphi^3 \over 32 \pi^3 \, v^2} 
 ~ \Big|  
    \sum_{\psi \, \in \, \psi_q}
      g_{\varphi \psi \psi} \, 
    f_{\rm \varphi}(\xi_{\psi} , \tau_{\psi}) \,  
    \Big|^2 
\eq
where $ \tau_{\psi} = 4 \, m_\psi^2 / m_\phi^2$ 
and $\xi_{\psi} =  \bar{m}^2_\psi  /  m_\psi^2  $
where $m_\psi$ is the pole mass for the kinematic loop factors and  
$\bar{m}_\psi$ is the $\overline{\rm MS}$ fermion mass for the coupling, in this case evaluated at a renormalization scale of 
$\mu = {\rm max}(m_\varphi, m_\psi)$. 
For pseudoscalar decay $C_\varpi = 1$ and for scalar $C_\phi= {4 \over 9}$. 
The scalar coupling loop functions for heavy and light fermions with $\tau_{\psi} >1 $ and $<1$ respectively 
are \cite{Gunion:1989we}
$$
f^{\rm heavy}_{\rm \phi}(\xi , \tau) =  \xi \, {3 \tau \over 2} \,  \left[ 1 + (1 - \tau) 
     \left[ \sin^{-1}(1 / \sqrt{\tau} ) \right]^2 \,  \right]
     ~~~~~~~~~~~~~~~~~
$$
\beq
f^{\rm light}_{\rm \phi}(\xi , \tau) = \xi \, {3 \tau \over 2} \, \bigg[ 1 - {1 \over 4}  (1 - \tau) 
     \bigg[  \ln \bigg(  { 1 + \sqrt{ 1 - \tau} \over 1 -\sqrt{ 1 - \tau}   }  \bigg)  \, - i \pi   \,        \bigg]^2  \, \, \bigg]
\eq
and likewise the pseudoscalar coupling loop functions are \cite{Gunion:1989we}
$$
f^{\rm heavy}_{\rm \varpi}(\xi , \tau) = \xi \, \tau \, 
     \left[\,  \sin^{-1}(1 / \sqrt{\tau} ) \right]^2 
        ~~~~~~~~~~~~~~~~~~~
$$
\beq
f^{\rm light}_{\rm \varpi}(\xi , \tau) =  - \ \xi \, { \tau \over 4}  \, 
     \bigg[  \ln \bigg(  { 1 + \sqrt{ 1 - \tau} \over 1 -\sqrt{ 1 - \tau}   }  \bigg)  \, - i \pi   \,        \bigg]^2  
\eq
At threshold $\tau_{\psi} \! = \! 1$ with $f^{\rm heavy}_{\phi}(1,1) \! = \!  f^{\rm light}_{\phi}(1,1) \! = \!  {3 \over 2}$ 
 and $f^{\rm heavy}_{\varpi}(1,1) = f^{\rm light}_{\varpi}(1,1) =  {\pi^2 /4}$. 
 In the heavy fermion limit  $f^{\rm heavy}_{\phi}(1,\infty) = f^{\rm heavy}_{\varpi}(1,\infty)      = 1$.

The decay width of a massive spin-zero boson $\varphi$ to a photon pair $\gamma \gamma$
arising at one-loop from a coherent sum over massive charged Dirac fermions $\psi$ with 
scalar couplings 
(\ref{scalarcouplings}) or pseudoscalar couplings (\ref{pseudoscalarcouplings})  
 for $\varphi = \phi$ or $ \varpi$ is 
 \beq
\Gamma( \varphi \to \gamma \gamma ) =  C_\varphi \, {\alpha^2 ~m_\varphi^3 \over 64 \pi^3 \, v^2} 
 ~ \Big|  
    \sum_{\psi}
      g_{\varphi \psi \psi} \, N_\psi \, Q_\psi^2 \, 
    f_{\rm \varphi}(\xi_{\psi} , \tau_{\psi}) \,  
    \Big|^2 
\eq
where $Q_\psi$ is the Dirac fermion charge and $N_\psi =1,3$ is the loop multiplicity for leptons, quarks. 
The dimensionless mass squared ratios $\tau_\psi$ and $\xi_\psi$, 
the overall coupling coefficient $C_\varphi$ and loop functions  $f_{\rm \varphi}(\xi_{\psi} , \tau_{\psi})$
are identical to those given above 
 for decay to a gluon pair. 
 





\newpage 

\begin{thebibliography}{100}





\bibitem{CMS:2019lwf}
CMS Collaboration, 
``Search for physics beyond the standard model in multilepton final states in proton-proton collisions at $\sqrt{s} =$ 13 TeV,''
JHEP \textbf{03}, 051 (2020)
doi:10.1007/JHEP03(2020)051
[arXiv:1911.04968 [hep-ex]].

\bibitem{CMS:2024ulc}
CMS Collaboration, 
``Search for a scalar or pseudoscalar dilepton resonance produced in association with a 
  massive vector boson or top quark-antiquark pair in multilepton events at $\sqrt{s}$ = 13{\,}{\,}TeV,''
Phys. Rev. D \textbf{110}, no.1, 012013 (2024)
doi:10.1103/PhysRevD.110.012013
[arXiv:2402.11098 [hep-ex]].



\bibitem{CMS:2021ctt}
CMS Collaboration, 
``Search for resonant and nonresonant new phenomena in high-mass dilepton final states at $ \sqrt{s} $ = 13 TeV,''
JHEP \textbf{07}, 208 (2021)
doi:10.1007/JHEP07(2021)208
[arXiv:2103.02708 [hep-ex]].



\bibitem{ATLAS:2019erb}
ATLAS Collaboration, 
``Search for high-mass dilepton resonances using 139 fb$^{-1}$ of $pp$ collision data 
collected at $\sqrt{s}=$13 TeV with the ATLAS detector,''
Phys. Lett. B \textbf{796}, 68-87 (2019)
doi:10.1016/j.physletb.2019.07.016
[arXiv:1903.06248 [hep-ex]].





\bibitem{CMS:2012fgd}
CMS Collaboration, 
``Search for a Light Pseudoscalar Higgs Boson in the Dimuon Decay Channel in $pp$ Collisions at $\sqrt{s}=7$ TeV,''
Phys. Rev. Lett. \textbf{109}, 121801 (2012)
doi:10.1103/PhysRevLett.109.121801
[arXiv:1206.6326 [hep-ex]].

\bibitem{CMS:2019buh}
CMS Collaboration, 
``Search for a Narrow Resonance Lighter than 200 GeV Decaying to a Pair of Muons in Proton-Proton Collisions at $\sqrt{s}$ =  TeV,''
Phys. Rev. Lett. \textbf{124}, no.13, 131802 (2020)
doi:10.1103/PhysRevLett.124.131802
[arXiv:1912.04776 [hep-ex]].


\bibitem{CMS:2021sch}
CMS Collaboration, 
``Search for long-lived particles decaying into muon pairs in proton-proton collisions at $ \sqrt{s} $ = 13 TeV 
collected with a dedicated high-rate data stream,''
JHEP \textbf{04}, 062 (2022)
doi:10.1007/JHEP04(2022)062
[arXiv:2112.13769 [hep-ex]].

\bibitem{CMS:2023hwl}
CMS Collaboration, 
``Search for direct production of GeV-scale resonances decaying to a pair of muons in proton-proton collisions at $ \sqrt{s} $ = 13 TeV,''
JHEP \textbf{12}, 070 (2023)
doi:10.1007/JHEP12(2023)070
[arXiv:2309.16003 [hep-ex]].




\bibitem{LHCb:2017trq}
LHCb Collaboration, 
``Search for Dark Photons Produced in 13 TeV $pp$ Collisions,''
Phys. Rev. Lett. \textbf{120}, no.6, 061801 (2018)
doi:10.1103/PhysRevLett.120.061801
[arXiv:1710.02867 [hep-ex]].

\bibitem{LHCb:2019vmc}
LHCb Collaboration, 
``Search for $A'\to\mu^+\mu^-$ Decays,''
Phys. Rev. Lett. \textbf{124}, no.4, 041801 (2020)
doi:10.1103/PhysRevLett.124.041801
[arXiv:1910.06926 [hep-ex]].

\bibitem{LHCb:2020ysn}
LHCb Collaboration, 
``Searches for low-mass dimuon resonances,''
JHEP \textbf{10}, 156 (2020)
doi:10.1007/JHEP10(2020)156
[arXiv:2007.03923 [hep-ex]].




\bibitem{CMS:2014ccx}
CMS Collaboration, 
``Search for neutral MSSM Higgs bosons decaying to a pair of tau leptons in $pp$ collisions,''
JHEP \textbf{10}, 160 (2014)
doi:10.1007/JHEP10(2014)160
[arXiv:1408.3316 [hep-ex]].

\bibitem{CMS:2015qnd}
CMS Collaboration, 
``Search for a Low-Mass Pseudoscalar Higgs Boson Produced in Association with a $b\bar{b}$ Pair in $pp$ Collisions at $\sqrt{s} =$ 8 TeV,''
Phys. Lett. B \textbf{758}, 296-320 (2016)
doi:10.1016/j.physletb.2016.05.003
[arXiv:1511.03610 [hep-ex]].

\bibitem{CMS:2018rmh}
CMS Collaboration, 
``Search for additional neutral MSSM Higgs bosons in the $\tau\tau$ final state in proton-proton collisions at $\sqrt{s}=$ 13 TeV,''
JHEP \textbf{09}, 007 (2018)
doi:10.1007/JHEP09(2018)007
[arXiv:1803.06553 [hep-ex]].

\bibitem{CMS:2022goy}
CMS Collaboration, 
``Searches for additional Higgs bosons and for vector leptoquarks in $\tau\tau$ final states in proton-proton collisions at $\sqrt{s}$ = 13 TeV,''
JHEP \textbf{07}, 073 (2023)
doi:10.1007/JHEP07(2023)073
[arXiv:2208.02717 [hep-ex]].




\bibitem{ATLAS:2014vhc}
ATLAS Collaboration, 
``Search for neutral Higgs bosons of the minimal supersymmetric standard model in $pp$ 
   collisions at $\sqrt{s}$ = 8 TeV with the ATLAS detector,''
JHEP \textbf{11}, 056 (2014)
doi:10.1007/JHEP11(2014)056
[arXiv:1409.6064 [hep-ex]].

\bibitem{ATLAS:2016ivh}
ATLAS Collaboration, 
``Search for Minimal Supersymmetric Standard Model Higgs bosons $H/A$ and for a $Z^{\prime}$ boson 
  in the $\tau \tau$ final state produced in $pp$ collisions at $\sqrt{s}=13$ TeV with the ATLAS Detector,''
Eur. Phys. J. C \textbf{76}, no.11, 585 (2016)
doi:10.1140/epjc/s10052-016-4400-6
[arXiv:1608.00890 [hep-ex]].

\bibitem{ATLAS:2017eiz}
ATLAS Collaboration, 
``Search for additional heavy neutral Higgs and gauge bosons in the ditau final state produced in 36 fb$^{-1}$ of 
$pp$ collisions at $ \sqrt{s}=13 $ TeV with the ATLAS detector,''
JHEP \textbf{01}, 055 (2018)
doi:10.1007/JHEP01(2018)055
[arXiv:1709.07242 [hep-ex]].

\bibitem{ATLAS:2020zms}
ATLAS Collaboration, 
``Search for heavy Higgs bosons decaying into two tau leptons with the ATLAS detector using $pp$ collisions at $\sqrt{s}=13$ TeV,''
Phys. Rev. Lett. \textbf{125}, no.5, 051801 (2020)
doi:10.1103/PhysRevLett.125.051801
[arXiv:2002.12223 [hep-ex]].

\bibitem{ATLAS:2024rzd}
ATLAS Collaboration, 
``Search for a light CP-odd Higgs boson decaying into a pair of {\ensuremath{\tau}}-leptons 
in proton-proton collisions at $ \sqrt{s} $ = 13 TeV with the ATLAS detector,''
JHEP \textbf{12}, 126 (2024)
doi:10.1007/JHEP12(2024)126
[arXiv:2409.20381 [hep-ex]].








\bibitem{ATLAS:2012lmu}
ATLAS Collaboration, 
``Measurement of Upsilon production in 7 TeV pp collisions at ATLAS,''
Phys. Rev. D \textbf{87}, no.5, 052004 (2013)
doi:10.1103/PhysRevD.87.052004
[arXiv:1211.7255 [hep-ex]].


\bibitem{ATLAS:2017prf}
ATLAS Collaboration, 
``Measurement of quarkonium production in proton{\textendash}lead and 
  proton{\textendash}proton collisions at $5.02~\mathrm {TeV}$ with the ATLAS detector,''
Eur. Phys. J. C \textbf{78}, no.3, 171 (2018)
doi:10.1140/epjc/s10052-018-5624-4
[arXiv:1709.03089 [nucl-ex]].





\bibitem{LHCb:2012aa}
LHCb Collaboration, 
``Measurement of Upsilon production in pp collisions at $\sqrt{s}$ = 7 TeV,''
Eur. Phys. J. C \textbf{72}, 2025 (2012)
doi:10.1140/epjc/s10052-012-2025-y
[arXiv:1202.6579 [hep-ex]].


\bibitem{LHCb:2018yzj}
LHCb Collaboration, 
``Measurement of $\Upsilon$ production in $pp$ collisions at $\sqrt{s}$ = 13 TeV,''
JHEP \textbf{07}, 134 (2018)
[erratum: JHEP \textbf{05}, 076 (2019)]
doi:10.1007/JHEP07(2018)134
[arXiv:1804.09214 [hep-ex]].




\bibitem{CMS:2013qur}
CMS Collaboration, 
``Measurement of the $\Upsilon(1S), \Upsilon(2S)$, and $\Upsilon(3S)$ Cross Sections in $pp$ Collisions at $\sqrt{s}$ = 7 TeV,''
Phys. Lett. B \textbf{727}, 101-125 (2013)
doi:10.1016/j.physletb.2013.10.033
[arXiv:1303.5900 [hep-ex]].

\bibitem{CMS:2015xqv}
CMS Collaboration, 
``Measurements of the $\Upsilon$(1S), $\Upsilon$(2S), and $\Upsilon$(3S) differential cross sections 
 in $pp$ collisions at $\sqrt{s} =$ 7 TeV,''
Phys. Lett. B \textbf{749}, 14-34 (2015)
doi:10.1016/j.physletb.2015.07.037
[arXiv:1501.07750 [hep-ex]].

\bibitem{CMS:2017dju}
CMS Collaboration,  
``Measurement of quarkonium production cross sections in $pp$ collisions at $\sqrt{s}=$ 13 TeV,''
Phys. Lett. B \textbf{780}, 251-272 (2018)
doi:10.1016/j.physletb.2018.02.033
[arXiv:1710.11002 [hep-ex]].

\bibitem{CMS:2026ccg}
CMS Collaboration, 
``Measurement of the $\Upsilon$(1S), $\Upsilon$(2S), and $\Upsilon$(3S) differential cross 
  sections in $pp$ collisions at $\sqrt{s}$ = 13.6 TeV,''
[arXiv:2601.20023 [hep-ex]].




\bibitem{Zhao:2025sna}
Y.~Zhao, J.~Liu, X.~Cheng, C.~Wang and Z.~Hu,
``Experimental Review of the Quarkonium Physics at the LHC,''
Symmetry \textbf{17}, no.9, 1521 (2025)
doi:10.3390/sym17091521
[arXiv:2509.10330 [hep-ex]].






\bibitem{CMS:2024zhe}
CMS Collaboration, 
``Enriching the physics program of the CMS experiment via data scouting and data parking,''
Phys. Rept. \textbf{1115}, 678-772 (2025)
doi:10.1016/j.physrep.2024.09.006
[arXiv:2403.16134 [hep-ex]].







\bibitem{CMS:2026mwx}
CMS Collaboration, 
``Search for low-mass resonances decaying to $\tau \tau$ and measurement of 
 the $\Upsilon \to \tau \tau$ decay in proton-proton collisions at $\sqrt{s}$ = 13.6 TeV,''
[arXiv:2605.25103 [hep-ex]].




\bibitem{Buckley:2026xcv}
M.~R.~Buckley, D.~Shih and I.~R.~Wang,
``Hiding in the Shadow of the Upsilon: Ditaus from a Light Pseudoscalar,''
[arXiv:2605.29289 [hep-ph]].




\bibitem{CLEO:2006uhx}
D.~Besson \textit{et al.} [CLEO],
``First Observation of $\Upsilon(3S) \to \tau^+ \tau^-$ and Tests of Lepton Universality in Upsilon Decays,''
Phys. Rev. Lett. \textbf{98}, 052002 (2007)
doi:10.1103/PhysRevLett.98.052002
[arXiv:hep-ex/0607019 [hep-ex]].


\bibitem{BaBar:2010esv}
P.~del Amo Sanchez \textit{et al.} [BaBar],
``Test of lepton universality in $\Upsilon(1S)$ decays at BaBar,''
Phys. Rev. Lett. \textbf{104}, 191801 (2010)
doi:10.1103/PhysRevLett.104.191801
[arXiv:1002.4358 [hep-ex]].

\bibitem{BaBar:2020nlq}
J.~P.~Lees \textit{et al.} [BaBar],
``Precision measurement of the ${\cal B}(\Upsilon(3S)\to\tau^+\tau^-)/{\cal B}(\Upsilon(3S)\to\mu^+\mu^-)$ ratio,''
Phys. Rev. Lett. \textbf{125}, 241801 (2020)
doi:10.1103/PhysRevLett.125.241801
[arXiv:2005.01230 [hep-ex]].




\bibitem{Frixione:2007vw}
S.~Frixione, P.~Nason and C.~Oleari,
``Matching NLO QCD computations with Parton Shower simulations: the POWHEG method,''
JHEP \textbf{11}, 070 (2007)
doi:10.1088/1126-6708/2007/11/070
[arXiv:0709.2092 [hep-ph]].

\bibitem{Alioli:2010xd}
S.~Alioli, P.~Nason, C.~Oleari and E.~Re,
``A general framework for implementing NLO calculations in shower Monte Carlo programs: the POWHEG BOX,''
JHEP \textbf{06}, 043 (2010)
doi:10.1007/JHEP06(2010)043
[arXiv:1002.2581 [hep-ph]].

\bibitem{Bagnaschi:2011tu}
E.~Bagnaschi, G.~Degrassi, P.~Slavich and A.~Vicini,
``Higgs production via gluon fusion in the POWHEG approach in the SM and in the MSSM,''
JHEP \textbf{02}, 088 (2012)
doi:10.1007/JHEP02(2012)088
[arXiv:1111.2854 [hep-ph]].

\bibitem{NNPDF:2017mvq}
R.~D.~Ball \textit{et al.} [NNPDF],
``Parton distributions from high-precision collider data,''
Eur. Phys. J. C \textbf{77}, no.10, 663 (2017)
doi:10.1140/epjc/s10052-017-5199-5
[arXiv:1706.00428 [hep-ph]].

\bibitem{Alwall:2014hca}
J.~Alwall, R.~Frederix, S.~Frixione, V.~Hirschi, F.~Maltoni, O.~Mattelaer, H.~S.~Shao, T.~Stelzer, P.~Torrielli and M.~Zaro,
``The automated computation of tree-level and next-to-leading order differential cross sections, and their matching to parton shower simulations,''
JHEP \textbf{07}, 079 (2014)
doi:10.1007/JHEP07(2014)079
[arXiv:1405.0301 [hep-ph]].

\bibitem{Alwall:2007fs}
J.~Alwall, S.~Hoche, F.~Krauss, N.~Lavesson, L.~Lonnblad, F.~Maltoni, M.~L.~Mangano, M.~Moretti, C.~G.~Papadopoulos and F.~Piccinini, \textit{et al.}
``Comparative study of various algorithms for the merging of parton showers and matrix elements in hadronic collisions,''
Eur. Phys. J. C \textbf{53}, 473-500 (2008)
doi:10.1140/epjc/s10052-007-0490-5
[arXiv:0706.2569 [hep-ph]].

\bibitem{NNPDF:2014otw}
R.~D.~Ball \textit{et al.} [NNPDF],
``Parton distributions for the LHC Run II,''
JHEP \textbf{04}, 040 (2015)
doi:10.1007/JHEP04(2015)040
[arXiv:1410.8849 [hep-ph]].

\bibitem{Frederix:2012ps}
R.~Frederix and S.~Frixione,
``Merging meets matching in MC@NLO,''
JHEP \textbf{12}, 061 (2012)
doi:10.1007/JHEP12(2012)061
[arXiv:1209.6215 [hep-ph]].


\bibitem{Bierlich:2022pfr}
C.~Bierlich, S.~Chakraborty, N.~Desai, L.~Gellersen, I.~Helenius, P.~Ilten, L.~L{\"o}nnblad, S.~Mrenna, S.~Prestel and C.~T.~Preuss, \textit{et al.}
``A comprehensive guide to the physics and usage of PYTHIA 8.3,''
SciPost Phys. Codeb. \textbf{2022}, 8 (2022)
doi:10.21468/SciPostPhysCodeb.8
[arXiv:2203.11601 [hep-ph]].






\bibitem{LEPWorkingGroupforHiggsbosonsearches:2003ing}
R.~Barate \textit{et al.} [LEP Working Group for Higgs boson searches, ALEPH, DELPHI, L3 and OPAL],
``Search for the standard model Higgs boson at LEP,''
Phys. Lett. B \textbf{565}, 61-75 (2003)
doi:10.1016/S0370-2693(03)00614-2
[arXiv:hep-ex/0306033 [hep-ex]].






\bibitem{Drozd:2014yla}
A.~Drozd, B.~Grzadkowski, J.~F.~Gunion and Y.~Jiang,
``Extending two-Higgs-doublet models by a singlet scalar field - the Case for Dark Matter,''
JHEP \textbf{11}, 105 (2014)
doi:10.1007/JHEP11(2014)105
[arXiv:1408.2106 [hep-ph]].




\bibitem{Argyropoulos:2022ezr}
S.~Argyropoulos and U.~Haisch,
``Benchmarking LHC searches for light 2HDM+a pseudoscalars,''
SciPost Phys. \textbf{13}, no.1, 007 (2022)
doi:10.21468/SciPostPhys.13.1.007
[arXiv:2202.12631 [hep-ph]].








\bibitem{Glashow:1976nt}
S.~L.~Glashow and S.~Weinberg,
``Natural Conservation Laws for Neutral Currents,''
Phys. Rev. D \textbf{15}, 1958 (1977)
doi:10.1103/PhysRevD.15.1958



\bibitem{Gunion:2002zf}
J.~F.~Gunion and H.~E.~Haber,
``The CP conserving two Higgs doublet model: The Approach to the decoupling limit,''
Phys. Rev. D \textbf{67}, 075019 (2003)
doi:10.1103/PhysRevD.67.075019
[arXiv:hep-ph/0207010 [hep-ph]].



\bibitem{Craig:2013hca}
N.~Craig, J.~Galloway and S.~Thomas,
``Searching for Signs of the Second Higgs Doublet,''
[arXiv:1305.2424 [hep-ph]].










\bibitem{Gunion:1989we}
J.~F.~Gunion, H.~E.~Haber, G.~L.~Kane and S.~Dawson,
``The Higgs Hunter's Guide,''
Front. Phys. \textbf{80}, 1-404 (2000)
doi:10.1201/9780429496448











\end{thebibliography}
\end{document}